\documentclass[aps,10pt,twocolumn,superscriptaddress,balancelastpage,showpacs,reprint]{revtex4-1}
\usepackage{graphicx}
\usepackage[caption=false]{subfig}
\usepackage{dcolumn}
\usepackage{bm}
\usepackage{mathtools}
\usepackage{color,soul}
\usepackage{multirow}
\usepackage{amsmath}
\usepackage{placeins}
\usepackage{physics}
\usepackage{tikz}

\begin{document}
\title{Non-Hermitian Quasi-Localization and Ring Attractor Neural Networks}
\date{\today}

\begin{abstract}
Eigenmodes of a broad class of ``sparse'' random matrices, with interactions concentrated near the diagonal, exponentially localize in space, as initially discovered in 1957 by Anderson for quantum systems.
Anderson localization plays ubiquitous roles in varieties of problems from electrons in solids to mechanical and optical systems.
However, its implications in neuroscience (where the connections can be strongly asymmetric) have been largely unexplored, mainly because synaptic connectivity matrices of neural systems are often ``dense'', which makes the eigenmodes spatially extended. Here, we explore roles that Anderson localization could be playing in neural networks by focusing on ``spatially structured'' disorder in synaptic connectivity matrices. Recently, neuroscientists have experimentally confirmed that the local excitation and global inhibition (LEGI) ring attractor model can functionally represent head direction cells in \emph{Drosophila melanogaster} central brain. We first study a non-Hermitian (i.e.~asymmetric) tight-binding model with disorder and then establish  a connection to the LEGI ring attractor model. We discover that (i) Principal eigenvectors of the LEGI ring attractor networks with structured nearest neighbor disorder are ``quasi-localized'', even with fully dense inhibitory connections. (ii) The quasi-localized eigenvectors play dominant roles in the early time neural dynamics, and the location of the principal quasi-localized eigenvectors predict an initial location of the ``bump of activity'' representing, say, a head direction of an insect. Our investigations open up a new venue for explorations at the intersection between the theory of Anderson localization and neural networks with spatially structured disorder.
\end{abstract}

\author{Hidenori Tanaka}
\affiliation{Department of Applied Physics, Stanford University, Stanford, CA 94305, USA}
\affiliation{School of Engineering and Applied Sciences and Kavli Institute for Bionano Science and Technology, Harvard University, Cambridge, MA 02138, USA}
\author{David R. Nelson}
\affiliation{Departments of Physics and Molecular and Cellular Biology, Harvard University, Cambridge, MA 02138, USA}

\maketitle

\section{Introduction}
A grand challenge in modern neuroscience is to understand how a population of neurons in the brain collectively integrate sensory information, perform computations, and give rise to behavior while respecting biological constraints. To understand this population dynamics of immense complexity, one approach is to experimentally construct a complete wiring diagram of neurons, the ``connectome'' \cite{seung2012connectome}.
However, extracting the full details of the connectome has been challenging. 
For example, the total number of neurons in a human brain is estimated to be $10^{11}$, and there are roughly $10^{15}$ connections between them \cite{sporns2005human} forming an extremely intricate web.

In the face of such complexity, random matrix theory has allowed theoretical insights capturing statistical aspects of the connections.
In 1988, Sompolinsky, Crisanti, and Sommers \cite{sompolinsky1988chaos} generalized Girko's circular law for eigenvalue distribution in the complex plane to explore the chaotic dynamics of randomly connected neural assemblies (as might be the case before pruning during early neural development) as a function of the variance of the synaptic connection strengths.
Later, Rajan and Abbott imposed Dale's law (connections from a given neuron are either excitatory or inhibitory) to take realistic constraints for biological neural networks into account \cite{rajan2006eigenvalue}.
In the above studies, elements of the random matrices describing connections between excitatory and inhibitory neurons are densely and uniformly distributed across the rows and columns without spatial structures.

Another way of approaching the problem is to study finely structured neural network models that can achieve specific tasks.
For example, direction-selective cells that collectively store scalar variable $\theta \in [0,2\pi]$ representing say, head direction, has inspired theoretical works on ring attractor networks.
The ring attractor neural networks have been theoretically studied intensively \cite{ben1995theory, hansel199813, Zhang1996, knierim2012attractor,xie2002double} including diffusion and drift of bump of activity due to disorder \cite{Zhang1996, renart2003robust, kilpatrick2013wandering, itskov2011short, zhong2018non}.
Strikingly, neurobiologists have recently discovered that head-direction cells are anatomically placed in a ring geometry in the \emph{Drosophila melanogaster} central brain and identified the network structure through systematic perturbations \cite{Kim2017}.
Key facts derived from a series of careful experiments and theoretical modeling \cite{Seelig, Kim2017, green2017neural, turner2017angular} are the following:
(i) In \emph{Drosophila melanogaster}'s central brain, neurons are anatomically placed on a ring topography and form the ellipsoid body.
(ii) There is a bump of neural activity on the ellipsoid body whose location around the ring represents the head-direction $\theta\in[0,2\pi]$.
(iii) Nearest-neighbor excitations of neurons on the ellipsoid body via E-PG and P-EN neurons are responsible for sustaining a stationary bump of activity.
(iv) Upon rotation of the head-direction, P-EN neurons create a nearest-neighbor asymmetric bias to conduct angular velocity integration.

In this article, inspired by the above discoveries, we focus on the interplay between spatial structure and randomness in structured neural networks.
The Anderson's theory of the localization of quantum mechanical eigenfunctions is perhaps the most famous example of this interplay \cite{anderson1958absence}.
Initially, Anderson introduced a random tight-binding model to study electron conduction of one-dimensional solids with impurities.
The striking discovery is that even small amount of disorder makes essentially all the eigenstates exponentially localized in space and turns conductors into insulators in one and two dimensions \cite{mott1995theory,lee1985disordered}.
This picture is now known to have ubiquitous implications for wide varieties of physical systems from ultracold atoms and optical systems to mechanical systems with random spring contacts \cite{roati2008anderson, billy2008direct, segev2013anderson, dyson1953dynamics, ishii1973localization}.
However, recently two works have addressed the implications for neuroscience\cite{chaudhuri2014diversity,Amir2016}. Motivated by the above theoretical insights and recent experimental developments in neuroscience, we ask ``What role can Anderson localization play for neural networks with a spatially structured disorder?''.

In the following, first, we define a fairly general class of a one-dimensional non-Hermitian random tight-binding models, with tunably asymmetric couplings between neighboring neurons \cite{Amir2016}, and investigate its complex eigenvalues and the localization properties.
We also apply an extension of the sparse matrix to model the ring attractor dynamics of head direction cells governed by dense, structured synaptic connectivity matrix. As mentioned above, neurobiologists have found that the head direction cells of \emph{Drosophila melanogaster} are topographically placed on a ring. A recent study conducted perturbation-response experiments and, by comparisons with theoretical predictions, argued that the network structure is local excitation with flat global inhibition \cite{Kim2017}. 
Remarkably, we have found for a simple model with a quenched random disorder in the excitatory connections a parameter range such that the principal (i.e., most rapidly growing) eigenvectors are nevertheless still ''quasi-localized'', even when we add uniform global inhibitory connections that make the synaptic connectivity matrix fully connected. 
We then demonstrate that such localized principal eigenvectors dominate the initial dynamics before saturation.

\section{Non-Hermitian localization in a random tight-binding model}
First, we define the one-dimensional non-Hermitian tight-binding matrix model $\boldsymbol{M}$ as
\begin{equation}
\begin{split}
\boldsymbol{M} &=   \sum_{i=1}^{N} \big( s_i^+ e^{+g}|i+1\rangle \langle i| + s_i^- e^{-g}|i\rangle \langle i+1| \big),\\
\end{split}
\label{defineM}
\end{equation}
with a periodic boundary condition $\ket{i+N} = \ket{i}$.
The ket vector $\ket{i}$ is the basis to represent the neural activity of a neuron $i$.
The strengths of each nearest-neighbor connection $s^{\pm}_i$ are independent and identically distributed (i.i.d.) random value drawn from double box probability distribution with width $u$, and the sign of a connection can be either excitatory or inhibitory,
\begin{equation}
P_s (s | u, f) =\left\{
  \begin{array}{@{}lll@{}}
    f/u, & \text{for}\ 1-u/2<s<1+u/2 \\
    (1-f)/u, & \text{for} -1-u/2<s<-1+u/2 \\
    0, & \text{otherwise.}
  \end{array}\right.
\label{probM}
\end{equation}
The parameter $f \in [0,1]$ controls the ratio of excitatory to inhibitory connections.
Although this model, strictly speaking, violates Dales law, it was shown in \cite{Amir2016} via a similarity transformation that its spectra are identical for large rank sparse random matrices to a closely related model that does obey this constraint.
Our parametrization is slightly different from a previously studied model \cite{Amir2016}, to avoid numerical instability after a similarity transformation. 
When $f=1$ and $0<u<1$, it is closely related to a model which was studied earlier to understand the statistical physics of vortex lines in superconductors with random columnar pins \cite{Hatano1997} and the spread of biological organisms through random environments \cite{Shnerb1998}.
More recent work \cite{Amir2016} has focused on the case of completely random sign connections ($f=1/2$, $u=0$) while varying the non-Hermitian asymmetry parameter $g$.
For $g \geq 0$, the magnitude of connections clockwise around the ring (excitatory or inhibitory) are systematically stronger than those in the counterclockwise direction.
In the following, we set the connection variance of the excitatory and inhibitory connections to be $u=0.5$ and investigate properties of the resulting one-dimensional sparse random matrices on the full $(f,g)$ plane. Thus, our investigation allows us to interpolate between these two previously studied limiting cases, including parameter regimes closely related to a ring attractor model with disorder.
\section{Localized eigenvectors and complex eigenvalue spectra}
\begin{figure} [!htb]
\centering
\includegraphics[clip,width=1.0\columnwidth]{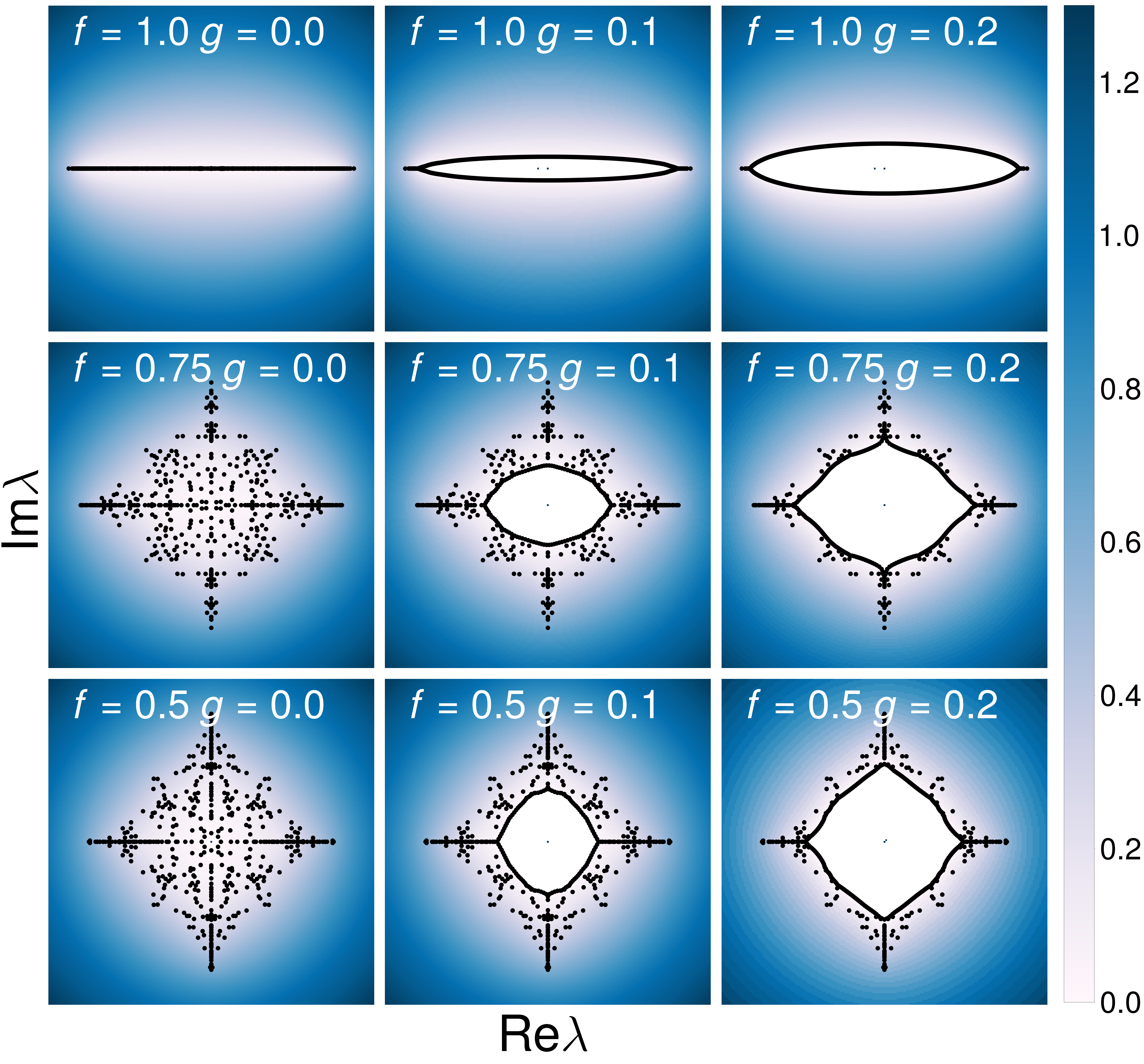}
\caption{
\textbf{Heat maps of the effective inverse eigenfunction localization lengths}\\
Heat maps of the effective inverse eigenfunction localization lengths $\kappa^{\rm{eff}}_\lambda= \frac{2 \kappa^+_\lambda \kappa^-_\lambda}{\kappa^+_\lambda + \kappa^-_\lambda}$ of one-dimensional sparse random matrices for various values of $(f,g)$. (See Appendix~\ref{Transfer} for the details behind our definition of $\kappa^{\rm{eff}}_{\lambda}$.)
The size of these sparse random matrices is $N=500$ and the strength of randomness of the matrix $\boldsymbol{M}$ in Eq.~(\ref{defineM}) is fixed at $u=0.5$, while the excitatory/inhibitory balance $f$ and clockwise bias $g$ are varied in $0.25$ and $0.1$ increments respectively.
No states are possible in the regions where $\kappa^{\rm{eff}}_\lambda<0$; these white regions correspond to energy gaps in the complex plane.
Eigenvalues obtained from numerical diagonalization of one particular realization are superimposed as black dots.
The inverse localization length $\kappa^{\rm{eff}}_\lambda$ is the largest, i.e., the effect of localization is the strongest, at the outer boundaries of these eigenvalue spectra, including the modes with the largest real parts that (as discussed below) dominate the neural dynamics.
See Fig.~\ref{Evec}(b) for representative principal eigenvectors that are strongly localized. As discussed in Ref.~\cite{Amir2016}, the localization length diverges near the origin for $g=0.0$, and as one approaches the rim of the hole (i.e., energy gap) in the spectra for $g>0$. These spectra are invariant under the transformation $g \rightarrow -g$ and $f \rightarrow (1-f)$.
}
\label{Hmap}
\end{figure}
We first display the complex eigenvalue spectra and the inverse localization length of the random tight binding model in the $(f,g)$ plane in Fig.~\ref{Hmap}.
We fixed the size of matrices to be $N=500$, randomness in magnitude to be $u=0.5$, and varied the excitatory/inhibitory ratio $f$ and the clockwise bias parameter $g>0$.
The row with $(f=0.5,g)$ was studied in Ref. \cite{Amir2016} without magnitude randomness ($u=0$) focusing purely on sign randomness (with $f=1/2$); Here we provide investigations of the properties of the matrix $\boldsymbol{M}$ in Eq.~(\ref{defineM}) over full parameter space on the $(f,g)$ plane with finite $u=0.5$.
We calculated the inverse localization length (in inverse lattice spacings, indicated by the heat map) by the transfer matrix method, as explained in the Appendix~\ref{Transfer}.
In Fig.~\ref{Hmap}, we see that the complex eigenvalue spectra observed for $f=0.5$ studied in \cite{Amir2016} are qualitatively robust up to $f=0.75$; in fact, we find similar spectra all the up to $f\approx 0.90$.
Importantly, the principal eigenvectors that play dominant roles in linear dynamics (i.e., those eigenvalues with the largest real parts) are most strongly localized ones, i.e., those with the shortest localization lengths $1/\kappa_{\lambda}^{\rm{eff}}$.
\section{Quasi-Anderson localization in local excitation/global inhibition (LEGI) ring attractor neural networks}
\begin{figure} [!htb]
\centering
\includegraphics[clip,width=1.0\columnwidth]{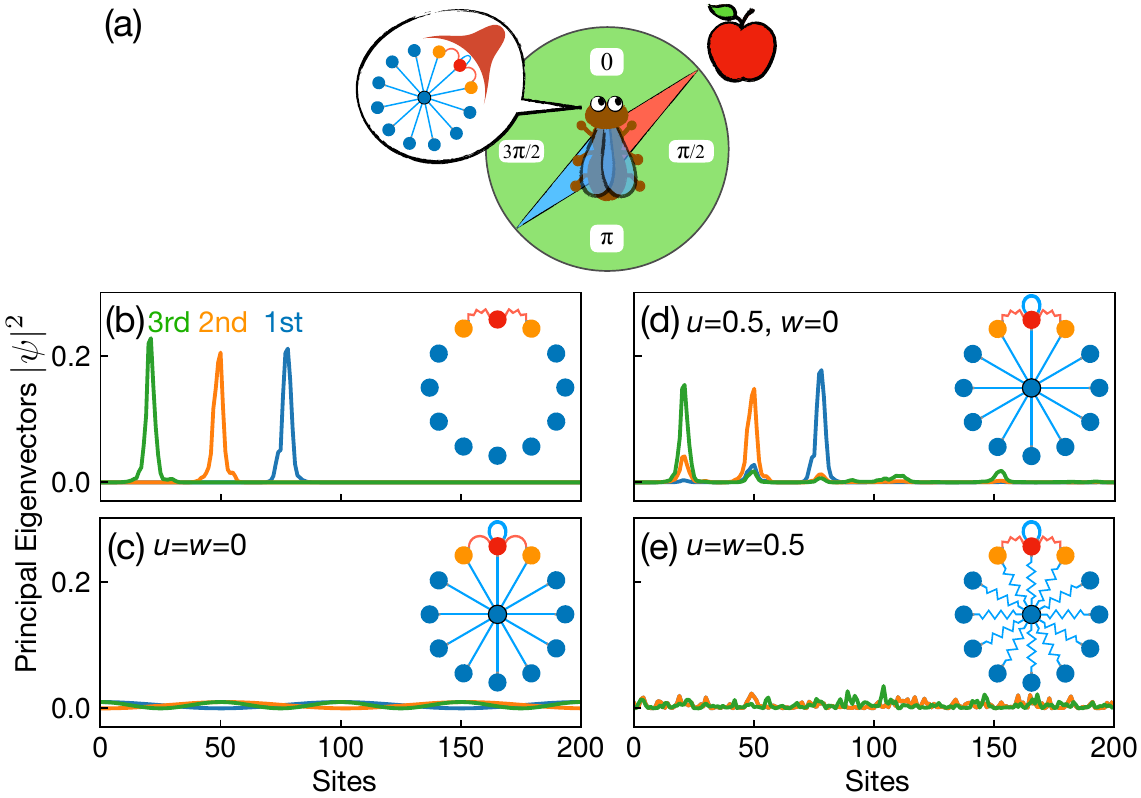}
\caption{
\textbf{Principal eigenvectors quasi-localize even with global inhibitory connections}\\
(a) ``Localized bump of activity'' (The blue lines feeding through the blue dot at the center represent a one-to-many inhibitory connection from each neuron on the ring to all other neurons) within a ring attractor neural network designed to model \emph{Drosophila melanogaster}'s central brain that represents the head direction $\theta\in[0,2\pi]$ \cite{Kim2017}.
(b) First, second, and third right principal eigenvectors (i.e. those whose eigenvectors have the largest real part) of the synaptic connectivity matrix $\boldsymbol{M}$ (Eq.~(\ref{defineM})) with excitatory/inhibitory balance $f=1.0$, clockwise bias $g=0$, and randomness $u=0.5$, for a ring with $N=200$ sites.
Each neuron has excitatory connections with random strengths to its two nearest neighbors (red lines in (b)).
The system is a strictly one-dimensional ring, and the principal eigenvectors are localized due to the random nearest-neighbor excitations with the same sign.
(c) Same three principal eigenvectors of a ring attractor neural network $\boldsymbol{J}$ without disorder, with a uniform set of inhibitory interactions (including self-inhibition) in blue. Without randomness, the synaptic connectivity $\boldsymbol{J}$ is a circulant matrix, and the eigenvectors are spatially extended around the ring in the form of sines and cosines.
(d) Principal eigenvectors of the ring attractor neural network $\boldsymbol{J'}$ with the same random nearest neighbor connections as in (a) and (b). Each neuron has two nearest-neighbor excitatory connections (red) with the same random strength ($u=0.5$) as in (b), with the addition of flat global inhibitory connections (blue) without randomness($w=0$).
With the flat global inhibitory connections, the synaptic connectivity matrix is fully dense. However, the principal eigenvectors are nevertheless ''quasi-localized'' with only small deviations from the eigenvectors shown in (a).
(e) The quasi-localization in (d) is destroyed when the randomness of the global inhibitory connections are as strong as the one of local excitatory connections, $u=w$.
}
\label{Evec}
\end{figure}
Here we establish a connection between $\boldsymbol{M}$ defined as in Eqs.~(\ref{defineM}),~(\ref{probM}) and a model of the head direction cells recently observed in the \emph{Drosophila melanogaster} central brain \cite{Kim2017, Turner-Evans2017}.
Computational neuroscientists have theoretically studied ring attractor networks to model head direction cells for many years \cite{ben1995theory, hansel199813, Zhang1996, knierim2012attractor,xie2002double}.
While experimental results have confirmed some of the predictions based on single neuron recordings, direct observation of the anatomical and functional network structure have been limited until quite recently \cite{heinze2017neural}.
In 2015, Seelig and Jayaraman \cite{Seelig} employed the two-photon excitation microscopy to discover that the head direction cells in the \emph{Drosophila} central brain are in fact anatomically placed with a ring topology and connectivity. (See Fig.~\ref{Evec}(a))
Very recently, this group further applied simultaneous optogenetic perturbation and two-photon calcium recordings to conclude that the properties of the recurrent neural network agrees well with theoretical predictions associated with a ``local excitation and global inhibition'' (LEGI) model \cite{Kim2017}. (see the inset of Fig.~\ref{Evec}(a).)\\
Mathematically, the synaptic connectivity matrix of the non-random LEGI model $\boldsymbol{J}$ can be written in a convenient bra-ket notation,
\begin{equation}
\boldsymbol{J}
=  \sum_{i=1}^{N} \big[ \gamma | i \rangle \langle i | + \alpha \big(e^{+g}|i+1\rangle \langle i| + e^{-g}|i\rangle \langle i+1| \big) \big]
 - \beta \sum_{j,k=1}^{N} |j\rangle \langle k|.
 \end{equation}
 The first-term represents self-excitation $\gamma$, the second term represents the nearest neighbor excitation $\alpha$, and the last term is the fully connected global inhibition $\beta$. We note that we added the factors of $e^{\pm g}$ to phenomenologically represent asymmetric bias for angular integration.
The continuous analogue of the above discrete model, after proper reparameterization to produce coupling constants $(\tilde{\alpha},\tilde{\beta},\tilde{\gamma}, \tilde{v})$ is (See Appendix~\ref{LEGImodel} for details) is,
\begin{equation}
\tau \partial_{t}r = -r + f\bigg[ \tilde{\gamma} r
- \tilde{v}(t) \partial_{\theta}r
+ \tilde{\alpha} \partial^2_{\theta} r - \tilde{\beta} \int_0^{2\pi} r d\theta
\bigg],
\end{equation} 
where $r(\theta,t)$ is the firing rate at time $t$ for a neuron at location $\theta$, $0\leq \theta <2\pi$, around the ring.
Next, to study the role of disorder in the synaptic connections, we introduce random variables $s^{\pm}_i$  for the local excitations, and $\beta'_{jk}$ for the global inhibitions respectively,
\begin{equation}
\begin{split}
\boldsymbol{J'} &= \sum_{i=1}^{N} \big[ \gamma | i \rangle \langle i | + \alpha \big(s^+_i e^{+g}|i+1\rangle \langle i| 
+ s^-_i e^{-g}|i\rangle \langle i+1| \big) \big]\\
& - \sum_{j,k=1}^{N} \beta'_{jk} |j\rangle \langle k|.
\label{Jtilde}
\end{split}
\end{equation}
The strengths of the disorder are characterized by $u,~w$ with the probability density functions as below,
\begin{equation}
P_s (s | u, f=1) =\left\{
  \begin{array}{@{}lll@{}}
    1/u, & \text{for}\ 1-u/2<s<1+u/2\\
    0, & \text{otherwise.}
  \end{array}\right.
\end{equation}
for $s=s^+$ or $s=s^-$, excitatory connections, and
\begin{equation}
P_{\beta} (\beta' | \beta, w) =\left\{
  \begin{array}{@{}lll@{}}
    1/w, & \text{for}\ \beta- w/2 < \beta' < \beta+ w/2 \\
    0, & \text{otherwise.}
  \end{array}\right.
\end{equation}
for the global inhibitory connections.
In Fig.~\ref{Evec}, we display the three principal eigenvectors corresponding to $g=0$ and (b) $\boldsymbol{M}(u=0.5)$, (c) $\boldsymbol{J}$ without randomness, (d) $\boldsymbol{J'}(u=0.5,w=0)$, and (e) $\boldsymbol{J'}(u=w=0.5)$.
These three eigenmodes correspond to eigenvalues with the three largest real parts.
The eigenvectors in Fig.~\ref{Evec} (b) are examples of non-Hermitian localization due to random excitatory nearest neighbor interactions, and the eigenvectors are exponentially localized at particular position around the ring.
Fig.~\ref{Evec} (c) presents the principal eigenvectors of a ring attractor neural network without randomness, $\boldsymbol{J}$. Since $\boldsymbol{J}$ is a circulant matrix without a disorder, the eigenvectors are sines and cosines and are spatially extended around the ring.
However, when we add disorder to the local excitation connections (with the same numerical values as in (b)), the matrix $\boldsymbol{J'}(u=0.5,w=0)$ has eigenvectors that are quasi-localized, exhibiting ``quasilocalized'' peaks in space as in Fig.~\ref{Evec} (d). This result is striking since the matrix is fully dense with global inhibitory connections. In Appendix~\ref{Qloc}, we study this quasi-localization in detail and study the effect of single anomalous hopping matrix element $\delta{\boldsymbol{M}} = \delta m (\ket{1}\bra{2} + \ket{2}\bra{1})$ on the spectrum. Finally, when the disorder is uniformly added to all the matrix elements without spatial structure $(u=w=0.5)$, the quasi-localization disappears as shown in Fig.~\ref{Evec} (e).
\section{Dynamics of a non-linear firing rate model}
\begin{figure} [!htb]
\centering
\includegraphics[clip,width=1.0\columnwidth]{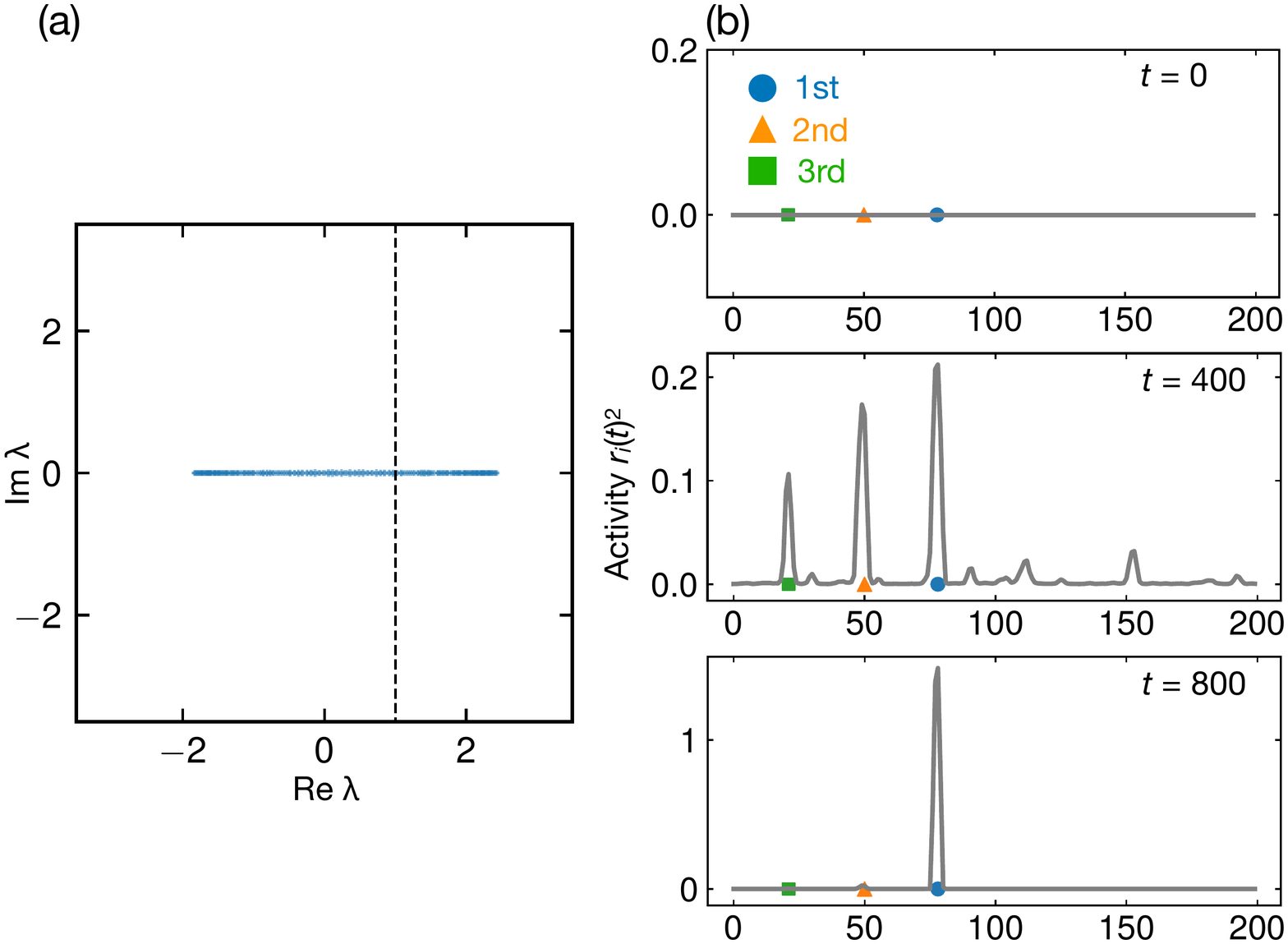}
\caption{
\textbf{Quasi-localized eigenvectors dominate the initial dynamics and bias the location of the stationary bump of activity}\\
(a) Eigenvalue spectrum of the LEGI model with their connectivity matrix $\boldsymbol{J'}$ with nearest neighbor randomness (see Eqs. (\ref{Jtilde})). All eigenvalues are real in this case. The vertical dashed line is at $\textnormal{Re}[ \lambda] =1$ and separates linearly stable eigenmodes from linearly unstable modes for the case of a nonlinear function given by $f[x] = (x+1)\Theta(x+1)$ in Eq.~\ref{Tevo}. For our choice of parameters (see below) more than just three principal eigenvectors will grow exponentially at short times from a state of negligible excitation.
(b) Time evolution of the activity of neurons $r_i (t)$ on the ring starting from an initial condition of homogeneous activity around the ring. Here, the neural ring has the short-range random excitation and long-range flat inhibition as shown in Fig.~\ref{Evec}(c).
The dynamics at the intermediate time ($t=400 \tau$) reflects the three principal eigenvectors plotted in Fig.~\ref{Evec}(d). The long-range inhibitory interactions have suppressed the remaining modes corresponding to $\lambda \geq1$ in (a). At later times, the system locks into a peak of activity dominated by the first principal localized mode (i.e., the eigenfunction with the largest real part).
The parameters used for this simulation are $N=200, \alpha=1, \beta=0.5, \gamma=0.3$, and $u=0.5$.
}
\label{dynamics}
\end{figure}

Here, we briefly review and study a firing-rate based non-linear recurrent neural network model incorporating the random matrices discussed above.
Biologically, neurons communicate by generating a series of electrochemical signals (action potentials).
A challenging aspect in mathematically modeling a discrete spike train is the coexistence of two separate time scales:
Each firing of a neural spike happens over roughly a millisecond.
In contrast, the characteristic timescale that governs emergent computations of the neural network (e.g., short-term memory) can be orders of magnitude longer!
A common strategy to incorporate this separation of time scales is to integrate the neural spike train over a time window to obtain the averaged firing rate denoted $r_i (t)$ for $i$-th neuron at time $t$.
Then the time evolution of $r_i (t)$ is governed by nonlinear dynamics which is first order in time,
\begin{equation}
\tau \frac{d r_i (t)}{dt} = - r_i (t) + f \bigg[ \sum_j J_{ij} r_j (t) + h_i (t) \bigg],
\label{Tevo}
\end{equation}
where the time scale is set by a reaction time $\tau$, $J_{ij}$ represents a strength of synaptic connection from neuron ``$j$'' to neuron ``$i$'', $f[\cdot]$ is nonlinear activation function that integrates over the inputs to neuron $i$, and $h_i (t)$ is an external input to this neuron from, say, the sensory system.
What are the implications of our study for the eigenvalue spectra and localized eigenvectors for the dynamics of the nonlinear neural network above when a synaptic connectivity matrix like $\boldsymbol{J'}$ in Eq.~(\ref{Jtilde}) is used?
Following \cite{Kim2017}, we assume neural activities are not saturating and use the ``threshold linear'' function $f(x) \equiv (x+1)\Theta (x+1) = [x+1]_+$ where $\Theta[y]$ is the Heaviside step function.
In Fig.~\ref{dynamics} (a), we plot a typical eigenvalue spectrum of $\boldsymbol{J'}(u=0.5,w=0)$ on complex plane.
Although the localized eigenvalue spectrum, in this case, lies on the real axis, we plot these in the complex eigenvalue plane to indicate the rich set of alternative dynamical models implicit in the more general spectra shown in Fig.~\ref{Hmap}.
The vertical dashed line at $\textnormal{Re} \lambda = 1$ separates growing and decaying modes, and we can confirm that for short times there are many more growing modes than the just those corresponding to the three principal eigenvectors.
In Fig.~\ref{dynamics} (b), we simulate the neural dynamics with the local excitation and global inhibition model $\boldsymbol{J'}$.
Starting from a flat initial condition, the plots show neural activities at $t=0,400,800$.
A plot at $t=400$ shows that just three principal eigenvectors dominate the initial dynamics before saturation.
In later times, most of the initial bumps down-regulate each other through the inhibitory connections: Only the first principal eigenvector survives to become a local bump of activity.
The location of the localized principal eigenvector successfully predicts the local bump of activity in the steady state, for a particular realization of the quenched random disorder.

\begin{figure} [!htb]
\centering
\includegraphics[clip,width=0.9\columnwidth]{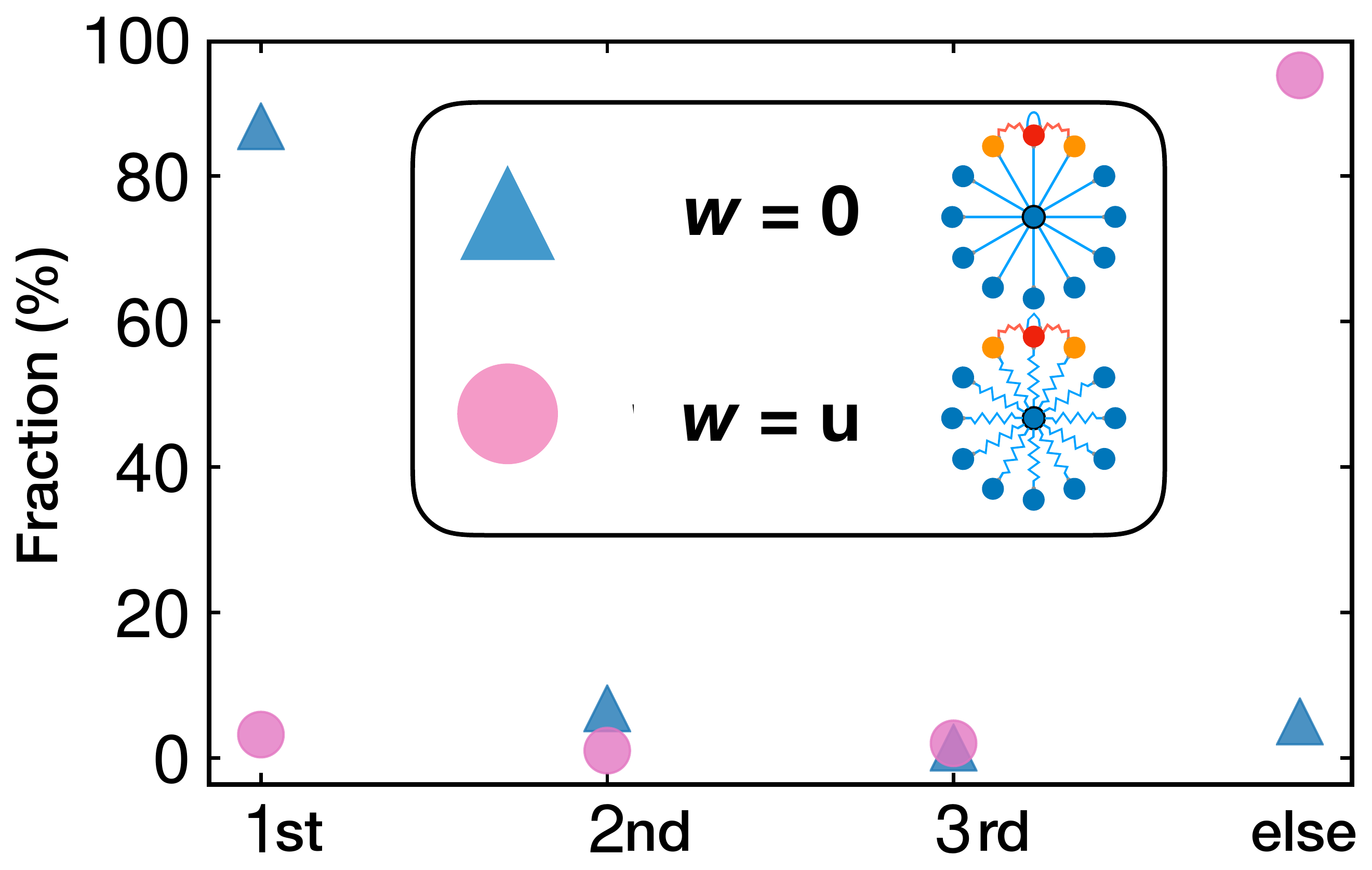}
\caption{\textbf{Histogram of the final locations of bumps of activity starting from homogeneous initial conditions}\\
Histogram of the final location of a bump of activity starting from a flat initial condition of activity around the ring.
The final location is classified as ''1st, 2nd, and 3rd'' when the distance to the peak of the localized principal eigenvectors is less than three sites away.
If the final location is far from any of these three localized positions, we classify this as ``elsewhere''.
When $u=0.5$, $w=0$ (in the case of Fig.~2 (d)), the quasi-localized first principal eigenvector location guides the final location of the localized bump of activity.
In particular, $87\%$ of the final locations are at the position of the first principal eigenvector. (see blue triangles)
However, such localization is absent when the disorder is uniformly distributed over the local excitations and global inhibitions with comparable strengths $u=w=0.5$.
In this case, only $5\%$ of the final locations of a local bump of activity are near a peak of the first principal eigenvector.
$94\%$ of the final positions are far from the peaks of any of three principal eigenvectors.
}
\label{histo}
\end{figure}

Next, we study the statistics of the final locations of the local bump of activity over 500 realizations of the disordered ring attractor network sampled from the probability distribution.
The histogram in Fig.~\ref{histo} shows how frequently the location of a bump of activity is near the site of the peaks of the three principal eigenvectors. When the network is in quasi-localized regime ($u=0.5$, $w=0$), the final locations are indeed mostly ($87\%$) near the peak of the first principal eigenvector.
However, once the eigenvectors are in the delocalized regime with uniform noise ($u=w=0.5$), only $5\%$ of the final locations are near the peak of the first principal eigenvectors.
\section{Selective excitation of the quasi-localized eigenmodes}
\begin{figure} [!htb]
\centering
\includegraphics[clip,width=1.0\columnwidth]{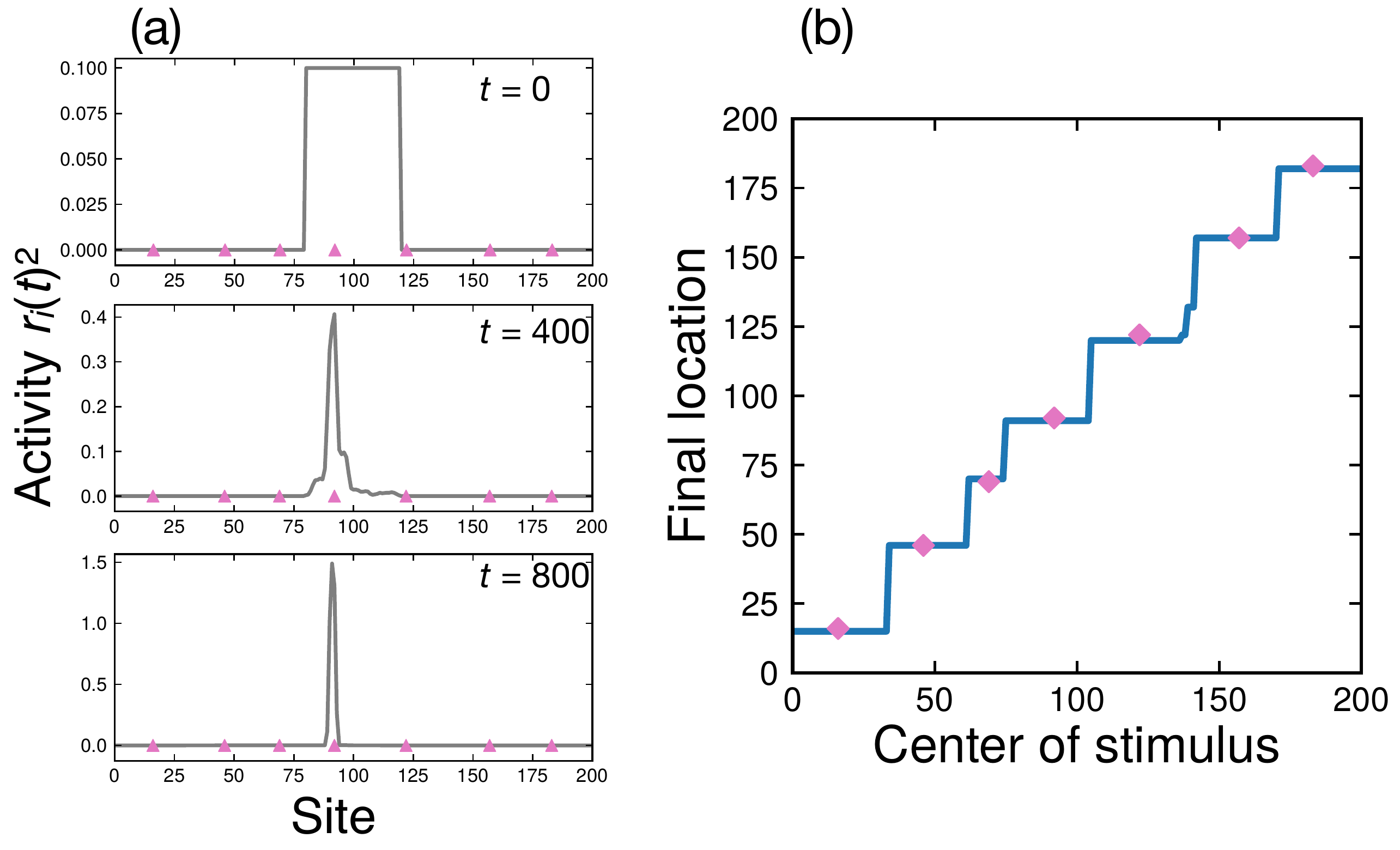}
\caption{
\textbf{Selective excitation of the quasi-localized principal eigenmodes from inhomogeneous initial conditions}\\
(a) The initial activity is set to $r_i(0)=0.1$ for $80 \leq i<120$ and otherwise $r_i (0) = 0$. All other parameters are the same as these in Fig.~\ref{dynamics}.
(b) [pink triangles] Locations (with periodic boundary conditions) of the quasi-localized eigenmodes with the seven largest real part of eigenvalues.
[blue line] The final position of the bump of activity as a function of a central location of a weak stimuli $c$. Each initial stimulus is applied with a $40$ site wide window ($c-20 \leq i \leq c+20$) as in (a).
The step-like structure of the line indicates that each of the discrete locations of the first seven localized eigenmodes can be selectively excited by broader initial external stimuli.
Because of the underlying disorder, we expect that these final locations will be insensitive to time-dependent noise, which would otherwise cause the bump of activity to diffuse around the ring.
}
\label{23dynamics}
\end{figure}

It is also of interest to selectively excite each of the principal eigenmodes by starting from inhomogeneous initial conditions as in Fig.~\ref{23dynamics}(a). Characteristics of the ring attractor network with quasi-localization may help to encode a discrete set of angular locations that is selectively retrievable. In Fig.~\ref{23dynamics}(b), we sweep the central location $c$ of the window of excitation ($c-20 \leq i \leq c+20$) and plot the final locations of the bump activity.
The step-like structure agrees well with the location of localized eigenmodes the with seven largest real part of eigenvalues.
With a particular realization of the quasi-localized eigenmodes, the disordered ring neural network can store seven discrete angles like a roulette wheel with irregular locations and a heavily overdamped frictional bearing.
\section{Discussion}
In summary, we first introduced a class of sparse non-Hermitian random matrix models that enabled us to systematically explore how randomness in sign and magnitude relates to eigenvalue spectra and spatial localization of eigenvectors (Fig.~1).
Then, we adapted the random matrix model to the observations of the dynamics of local excitation and global inhibition (LEGI) neural network model that was recently proposed to explain angular representation and integration in \emph{Drosophila melanogaster}'s central brain.
Rather surprisingly, we discovered that the principal eigenvectors are quasi-localized even with global inhibition ($\beta$) that makes the synaptic connectivity matrix fully dense (Fig.~2), provided the noise on the inhibition connections are small.
The quasi-localized eigenvector with the largest real eigenvalues dominates the initial dynamics and indicates the final location of the bump of activity upon starting from a uniform initial condition (Fig.~3,4).
Furthermore, by starting from a broad angular window of initial stimuli, we can selectively excite discrete localized modes.
The ring neural network with random excitatory nearest neighbor interactions and long-range inhibition can store long-term memory of discrete angles like a roulette wheel with somewhat irregular locations (Fig.~5).
Thus, the quasi-localized eigenvectors can store long-term memory of points in angle/space by weakly modulating structured matrix, while maintaining the functional properties. Furthermore, the localized spots reside in distinct eigenmodes associated with different eigenvalues.
Thus, each of the memorized points in angle/space can be separately excited by external stimuli.
It is an open problem to develop synaptic learning rules that can achieve precise control of the locations of the localized modes.
This work provides an initial step towards exploring the role of Anderson localization in neuroscience where a much richer variety of phenomena are possible due to the presence of both excitatory \& inhibitory interactions;
In the future, explorations at the intersection between biological networks and localization phenomena of random matrices with a spatially structured disorder, including low-rank matrices~\cite{mastrogiuseppe2017linking}, would be of interest.

\begin{acknowledgements}
We'd like to acknowledge inspiring and critical discussions with Shaul Druckmann, Richard Hahnloser, Ann Hermundstad, Vivek Jayaraman, Daniel D. Lee, Haim Sompolinsky, Misha Tsodyks.
This work was supported by the NSF, primarily through grant No.~DMR-1608501 and also the through the Harvard Materials Research Science and Engineering Center, via Grant No.~DMR-1420570.
\end{acknowledgements}

\appendix

\section{Transfer matrix method and the inverse localization length}
\label{Transfer}
Here we review a method to obtain the ``inverse localization length'' by using transfer matrix approach following Refs. \cite{Amir2016,furstenberg1960products}.
An eigenvector $\ket{\psi}$ of the tridiagonal matrix $\boldsymbol{M}$ and an eigenvalue $\lambda$ satisfies, 
\begin{equation}
\boldsymbol{M}\ket{\psi} = \lambda \ket{\psi},
\end{equation}
whose $n$-th element leads to
\begin{equation}
\textnormal{\bf{M}}_{n,n+1} \psi_{n+1} + \textnormal{\bf{M}}_{n,n} \psi_{n} + \textnormal{\bf{M}}_{n,n-1} \psi_{n-1} = \lambda \psi_{n},
\end{equation}
and thus, $\psi_{n+1}$ can be recursively related to $\psi_n$ and $\psi_{n-1}$ via
\begin{equation}
\psi_{n+1} =  \frac{\lambda - \textnormal{\bf{M}}_{n,n}}{\textnormal{\bf{M}}_{n,n+1}} \psi_{n} -  \frac{\textnormal{\bf{M}}_{n,n-1} }{\textnormal{\bf{M}}_{n,n+1} } \psi_{n-1}.
\label{recurs}
\end{equation}
This relation implies that, given a random matrix $\textnormal{\bf{M}}$ and a (possibly complex) eigenvalue $\lambda$, we can iteratively calculate $\psi_{n}$ $(n>2)$ from the first two elements of the eigenvector $\psi_{1}$ and $\psi_{2}$. 
We can further simplify the relation above and avoid numerical problem associated with the exponential divergence of $|\psi|$ by introducing a ratio between the neighbor elements (the ``Ricatti variable'')
\begin{equation}
r_{n}=\frac{\psi_{n+1}}{\psi_{n}}.
\end{equation}
The recursive relation (\ref{recurs}) then becomes
\begin{equation}
r_{n} =  \frac{\lambda - \textnormal{\bf{M}}_{n,n}}{\textnormal{\bf{M}}_{n,n+1}}  -  \frac{\textnormal{\bf{M}}_{n,n-1} }{\textnormal{\bf{M}}_{n,n+1} r_{n-1} }.
\end{equation}

The more specific form of the recursion relation for the random matrix of interest in this paper is
\begin{equation}
r_{n} =  \frac{\lambda e^{-g}}{s^+_n }  -  \frac{ s^-_{n-1} e^{-2g}}{ s^+_{n} r_{n-1} }.
\end{equation}

It follows from the definition of the Ricatti variable that
\begin{equation}
\begin{split}
|\psi_N| \approx |\psi_1| | \frac{\psi_2}{\psi_1} | | \frac{\psi_3}{\psi_2} | ... | \frac{\psi_N}{\psi_{N-1}} |
= |\psi_1| \prod_{i=1}^{N-1} r_i,
\end{split}
\end{equation}
and, assuming exponential localization $\frac{|\psi_N|}{|\psi_1|} = e^{\kappa_\lambda (N-1)}$ for a particular eigenvalue $\lambda$, we have
\begin{equation}
\begin{split}
\frac{|\psi_N|}{|\psi_1|} = e^{\kappa_\lambda (N-1)}=\prod_{i=1}^{N-1} r_i,
\label{expolocali}
\end{split}
\end{equation}
where $\kappa_\lambda$ is the inverse localization length.

Using this relation, we can compute $\kappa^{+}_\lambda$as
\begin{equation}
\kappa^{+}_\lambda=\frac{1}{N-1}\sum\limits_{i=1}^{N-1} \ln |r_i|.
\end{equation}

For non-Hermitian random matrix, two tails falling off from the opposite side of a localized wave function have different exponent due to the directional asymmetry ($e^{\pm g}$) induced in hopping terms.   
We can repeat the same calculation from the opposite side ($\psi_N$),
\begin{equation}
\frac{1}{r_{n-1}} = \frac{\lambda e^{+g}}{s^-_{n-1}} - \frac{s^+_n e^{+2g}}{s^-_{n-1}}r_n.
\end{equation}
In this case, $\kappa^-_\lambda$ is given by
\begin{equation}
\kappa^{-}_\lambda=\frac{1}{N-1} \ln \frac{|\psi_1|}{|\psi_N|} =\frac{1}{N-1}\sum\limits_{i=1}^{N-1} \ln \frac{1}{|r_i|}.
\end{equation}
We can now define an effective inverse localization length via the participation ratio \cite{Amir2016} for these asymmetrical wave functions via,
\begin{equation}
\kappa_{\lambda}^{\text{eff}} \equiv \frac{\sum_j |\psi_j|^4}{\sum_j |\psi_j|^2} = \frac{2\kappa_+ \kappa_-}{\kappa_+ + \kappa_-}
\label{InvPR}
\end{equation}
where we have inserted the exponential behavior embodied in, e.g. Eq.~(\ref{expolocali}).
The $\kappa_{\lambda}^{\text{eff}}=\frac{2\kappa_+ \kappa_-}{\kappa_+ + \kappa_-}$ is the effective inverse localization length used in Fig.~\ref{Hmap}.

\section{Local excitation and global inhibition (LEGI) model}
\label{LEGImodel}
\subsection{Continuous model}
Here, we start with the continuous local excitation and global inhibition (LEGI) model introduced in \cite{Kim2017}, and then add an asymmetric bias term $\tilde{v}\partial_{\theta} r$ that phenomenologically represents excitatory inputs from external neurons governing angular velocity integration \cite{skaggs1995model}. Recently, such a mechanism has been observed experimentally in P-EN neurons of \emph{Drosophila melanogaster} central brain \cite{green2017neural, turner2017angular}.
The integro-differential equation of our interest is
\begin{equation}
\begin{split}
&\tau \frac{\partial r(\theta,t)}{\partial t} = -r(\theta,t) + f\bigg[ \tilde{\gamma} r(\theta,t) 
- \tilde{v}(t) \frac{\partial}{\partial \theta} r(\theta,t)\\
&+ \tilde{\alpha} \frac{\partial^2 r(\theta,t)}{\partial \theta^2} - \tilde{\beta} \int_0^{2\pi} r(\theta,t) d\theta
\bigg],
\label{DefCont}
\end{split}
\end{equation}
where $\tau$ is a relaxation time for the neural firing rate, $r(\theta,t)$ is an activity of a neuron at angle $\theta$ at time $t$, $\tilde{\gamma}$ is the self excitation/inhibition, $\tilde{v}(\theta,t)$ is the strength of the asymmetric bias~\cite{skaggs1995model, green2017neural, turner2017angular}, $\tilde{\alpha}$ is the strength of the local excitatory connection, and $\tilde{\beta}$ represents the global inhibition.
\subsection{Discretization of the continuous LEGI model}
We now discretize each of above terms with $N$ neurons by writing
\begin{equation}
\begin{split}
&\frac{\partial r}{\partial \theta} \approx \frac{ r_{n+1}(t) - r_{n-1}(t)}{2 \Delta \theta },~
\frac{\partial^2 r}{\partial \theta^2} \approx \frac{ r_{n+1}(t) + r_{n-1}(t) - 2r_n (t)}{ \Delta \theta^2 },\\
&\int_0^{2\pi} r(\theta, t) d \theta \approx  \Delta \theta \sum_{m=1}^{N} r_m (t),
\end{split}
\end{equation}
where we define $r_n (t)$ and $\Delta \theta$ as
\begin{equation}
r_n (t) \equiv r \big(n\Delta \theta, t \big),~\Delta \theta = \frac{2\pi}{N}.
\end{equation}
Thus, we obtain the discretized version of the LEGI model with proper scaling,
\begin{equation}
\begin{split}
&\tau \frac{d r_n}{dt} = - r_n + f\bigg[ \tilde{\gamma} r_n -\tilde{v}(t) \bigg(\frac{N}{2\pi}\bigg) \frac{r_{n+1}-r_{n-1}}{2} \\
&+ \tilde{\alpha} \bigg( \frac{N}{2\pi} \bigg)^2 (r_{n+1}+r_{n-1}-2r_n) - \tilde{\beta} \bigg( \frac{2\pi}{N} \bigg) \sum_{m=1}^{N} r_m \bigg].
\end{split}
\end{equation}
Thus, $(\tilde{\alpha}, \tilde{\beta}, \tilde{\gamma})$ appearing in the continuous model Eq.~(\ref{DefCont}) can be rescaled such that $\alpha = \tilde{\alpha} \big( \frac{N}{2\pi}\big)^2$, $\beta = \tilde{\beta} \big( \frac{2\pi}{N} \big)$, $\gamma = \tilde{\gamma}-2\tilde{\alpha} \big( \frac{N}{2\pi} \big)^2 $, and $v = \tilde{v}(t) \big( \frac{N}{2\pi}\big)$ as a function of the total number of neurons $N$.
With the re-parameterized variables $(\alpha, \beta, \gamma)$, the discrete model reads
\begin{equation}
\begin{split}
&\tau \frac{d r_n}{dt} = - r_n\\
&+f\bigg[ \gamma r_n -v(t) \frac{r_{n+1}-r_{n-1}}{2} + \alpha (r_{n+1}+r_{n-1}) - \beta \sum_{m=1}^{N} r_m \bigg],
\end{split}
\end{equation}
where $f[x]$ is a nonlinear activation function, similar to Eq.~(8) defined in the main text.
Equivalently, we can summarize the neural dynamics as
\begin{equation}
\tau\frac{d \boldsymbol{r}}{dt} = - \boldsymbol{r} +f(\boldsymbol{J}\boldsymbol{r}),
\end{equation}
where the synaptic connectivity matrix, in a convenient bra-ket notation is
\begin{equation}
\begin{split}
&\boldsymbol{J}=\sum_{i=1}^{N} \big[ \gamma |i\rangle \langle i| \\
&+ \alpha \big( (1+v/2\alpha)|i+1\rangle \langle i| + (1-v/2\alpha)|i\rangle \langle i+1| \big) \big] - \beta \sum_{j,k=1}^{N} |j\rangle \langle k|\\
& \approx \sum_{i=1}^{N} \big[ \gamma |i\rangle \langle i| + \alpha \big( e^{+g} |i+1\rangle \langle i| + e^{-g} |i\rangle \langle i+1| \big) \big]
 - \beta \sum_{j,k=1}^{N} |j\rangle \langle k|,
\end{split}
\end{equation}
where the exponentiated hopping bias parameter $g$ is defined as $g=v/2\alpha$.
The calculations in this paper were done with a threshold linear activation function $f(x)\equiv (x+1) \Theta [x+1] = [x+1]_{+}$

\section{Study of quasi-localization}
\label{Qloc}
\subsection{Minimal model to study ``quasi-localization''}
To focus on the study of quasi-localized principal eigenmodes, here we focus on the simplified model shown below,
\begin{equation}
\boldsymbol{M} =  \sum_{i=1}^{N} \alpha \big( s_i^+ e^{+g}|i+1\rangle \langle i| + s_i^- e^{-g}|i\rangle \langle i+1| \big)
 - \sum_{j,k=1}^{N} \beta |j\rangle \langle k|,
\label{Ltilde}
\end{equation}
where the first two terms represent nearest neighbor couplings with an asymmetric nearest-neighbor bias parameter $g$, and $\beta$ describes global inhibition around the ring.
Note that the matrix $\boldsymbol{M}$ defined in the main text corresponds to a limited regime of $\beta=0$ in the above definition. For simplicity, we keep the same notation $\textbf{M}$ here. 
Upon taking $N=5$ for illustration purposes, we have
\begin{equation}
\begin{split}
&\boldsymbol{M}=\\
&\begin{pmatrix}
    0 &  s^-_{1} e^{-g} & 0 & 0 & s^+_{5} e^{+g} \\
    s^+_{1} e^{+g} & 0 & s^-_{2} e^{-g} & 0 & 0 \\
    0 & s^+_{2} e^{+g} & 0 & s^-_{3} e^{-g} & 0 \\
    0 & 0 & s^+_{3} e^{+g} & 0 & s^-_{4} e^{-g} \\
    s^-_{5} e^{-g} & 0 & 0 & s^+_{4} e^{+g} & 0 
  \end{pmatrix}
  -
    \begin{pmatrix}
    \beta & \beta & \beta & \beta & \beta\\
    \beta & \beta & \beta & \beta & \beta\\
    \beta & \beta & \beta & \beta & \beta\\
    \beta & \beta & \beta & \beta & \beta\\
    \beta & \beta & \beta & \beta & \beta
  \end{pmatrix}.
\end{split}
\end{equation}
In contrast to the matrix $\boldsymbol{J}$, we ignore the diagonal elements that simply shift all the eigenvalues by a constant while leaving the eigenvectors unchanged. Specifically, if a matrix has a right eigenvector $\bm{A}\ket{\psi_n} = a_n \ket{\psi_n}$, constant diagonal elements $\gamma$, simply shift the entire eigenvalue $a_n$ acending to
\begin{equation}
(\bm{A} + \gamma \bm{I})\ket{\psi_n} = (a_n+\gamma) \ket{\psi_n}.
\end{equation}
We characterize the disorder in $s^{\pm}_i$ by parameter $u$ appearing in the probability density functions as below,
\begin{equation}
P_s (s | u, f) =\left\{
  \begin{array}{@{}lll@{}}
    f/u, & \text{for}\ 1-u/2<s<1+u/2\\
    (1-f)/u, & \text{for} -1-u/2<s<-1+u/2\\
    0, & \text{otherwise.}
  \end{array}\right.
\end{equation}
\subsection{Localization properties}
\begin{figure} [!b]
\includegraphics[clip,width=0.9\columnwidth]{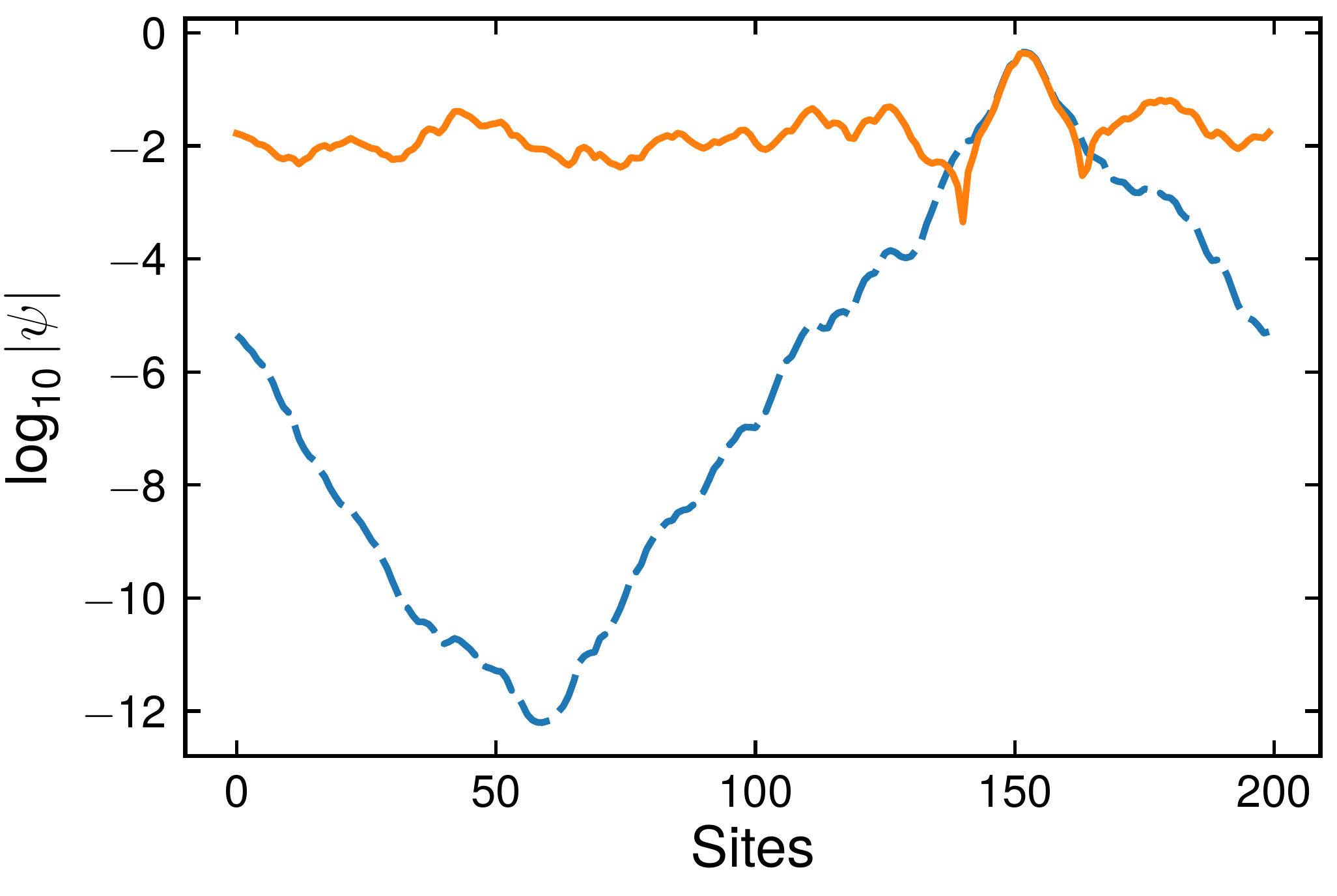}
\caption{The principal eigenvectors with largest (real) eigenvalue, with $N=200$, $f=1$, $u=0.5$ on a log-scale, and the periodic boundary conditions appropriate to a ring.
The blue dashed line is the conventional exponential localization without global inhibition ($\beta=0$) and the orange solid line corresponds to the quasi-localized eigenmode with global inhibition ($\beta=1$), with the same realization of the disordered nearest neighbor interactions. Note that we plot the logarithm of the modulus of the dominant eigenfuction and both eigenfunctions are strongly peaked near lattice site $150$.}
\label{logscale}
\end{figure}

\begin{figure*} [!b]
\includegraphics[clip,width=0.80\paperwidth]{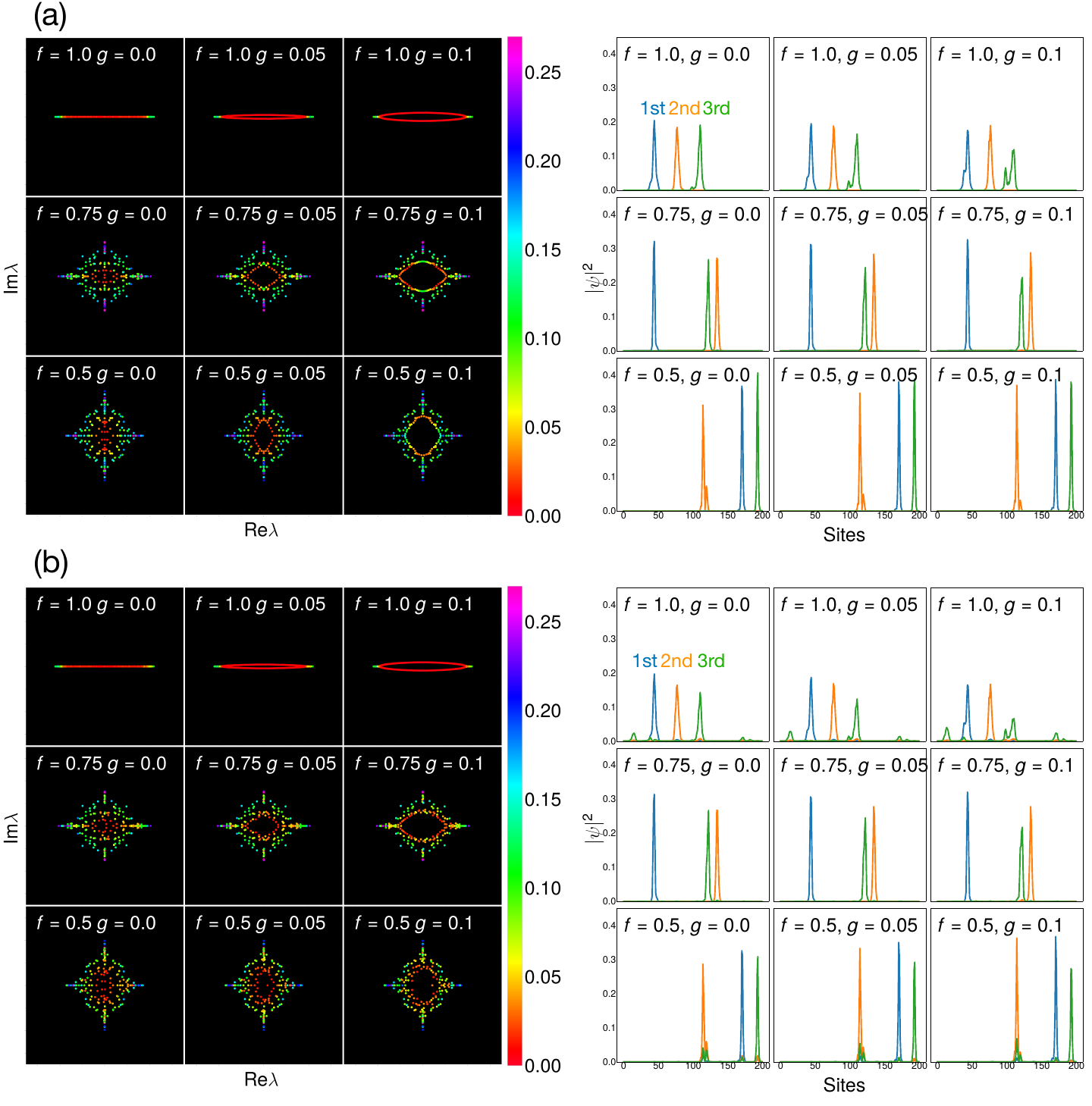}
\caption{(a) [Left] Complex eigenvalue spectra with varying ratio $f$ of excitatory to inhibitory interactions and asymmetry parameter $g$. These spectra are colored based on the inverse participation ratio without global inhibition ($\beta=0$).
[Right] The modulus of the three principal eigenvectors corresponding to each of the parameter set. The eigenvectors with the largest $\text{Re} \lambda$ are shown in green (these are the most strongly localized) while these with the next two largest $\text{Re} \lambda$ are shown in blue and red respectively. The eigenvectors are exponentially localized in space.
(b) [Left] Similar to (a), we show complex eigenvalue spectra colored based on the inverse participation ratio with global inhibition ($\beta=1$) without randomness ($w=0$).
[Right] Three principal eigenvectors corresponding to each of the parameter regime. The eigenvectors are still quasi-localized even with the global inhibition.
}
\label{Figure7}
\end{figure*}

Here, we present a few representative plots of the complex eigenvalue spectra and the localized eigenmodes.
Fig.~\ref{logscale} shows a semi-log plot of the principal eigenvector of the matrix $\boldsymbol{M}$ with (orange solid line, $\beta=1$) and without (blue dashed line, $\beta=0$) global inhibitory connections.
While the orange solid line is not exponentially localized in a strict sense, the amplitude $|\psi|$ nevertheless decays exponentially until it is down by a factor of more than $10^3$, $\sim$ ten lattice sites away from its center of localization.
In Fig.~\ref{Figure7} we present complex eigenvalue spectra and corresponding three principal eigenvectors.
The eigenvalue spectra is colored according to the inverse localization length obtained by calculating the inverse participation ratio $\kappa^{\text{eff}} = \frac{\sum\limits_j |\psi_j|^4}{\sum\limits_i |\psi_i|^2}$.
\clearpage
\onecolumngrid
\clearpage
\twocolumngrid

\subsection{Bloch/Floquet eigenfunctions of the LEGI model without disorder}
We now discuss the Bloch/Floquet eigenfunctions that result when all the nearest neighbor interactions are excitatory ($f=1$), and there is no disorder ($u=0$) upon setting the interaction magnitude $\alpha=1$, $\boldsymbol{M}$ is simply a function of $g$, and $\beta$,
\begin{equation}
\boldsymbol{M}(g) = \sum\limits_{m,n=1}^{N} \big[ (e^{+g} \delta_{m,n+1} + e^{-g} \delta_{m+1,n}) - \beta \big] \ket{m}\bra{n},
\label{Ltilde}
\end{equation}
or in matrix notation (taking $N=5$ for concreteness)
\[
\boldsymbol{M}(g)=
  \begin{pmatrix}
    -\beta & e^{-g}-\beta & -\beta & -\beta & e^{+g}-\beta \\
    e^{+g}-\beta & -\beta & e^{-g}-\beta & -\beta & -\beta \\
    -\beta & e^{+g}-\beta & -\beta & e^{-g}-\beta & -\beta \\
    -\beta & -\beta & e^{+g}-\beta & -\beta & e^{-g}-\beta \\
    e^{-g}-\beta & -\beta & -\beta & e^{+g}-\beta & -\beta 
  \end{pmatrix}.
\]
Since $\boldsymbol{M}$ is a circulant matrix without disorder, it can be diagonalized by the (delocalized) orthonormal Bloch/Floquet states, indexed by $k=\frac{2\pi}{N}s$, $s=0,\pm1,...,$
\begin{equation}
\tilde{\ket{k}} \equiv \frac{1}{\sqrt{N}} \sum_{l=1}^{N} e^{ikl} \ket{l},
~\langle\tilde{k}|\tilde{k'}\rangle = \frac{1}{N} \sum_{l=1}^{N} e^{i(k'-k)l} = \delta_{k,k'}.
\end{equation}
The corresponding eigenvalues can be calculated as follows,
\begin{equation}
\begin{split}
&\boldsymbol{M} ( \sqrt{N} \tilde{\ket{k}} ) \\
&=\sum\limits_{m,n=1}^{N} \Big[ \big(e^{+g} \delta_{m,n+1} + e^{-g} \delta_{m+1,n}\big) - \beta \Big] \ket{m}\bra{n} \big( \sum_{l=1}^{N} e^{ikl} \ket{l}  \big)\\
&= \sum_{m=1}^{N} \Big[ \big( e^{-i(k+ig)} + e^{i(k+ig)} \big) e^{ikm} - \beta\big( \sum_{n=1}^{N} e^{ikn} \big) \Big] \ket{m}\\
&=  \Big[ 2\cos(k+ig) - \beta N \delta_{k,0} \Big] \bigg( \sum_{m=1}^{N} e^{ikm} \ket{m}\bigg),\\
&\equiv E(k) ( \sqrt{N} \tilde{\ket{k}} ),
 \end{split}
\end{equation}
where the eigenvalue $E(k)$ is,
\begin{equation}
E(k) = 2\cos(k+ig) - \beta N \delta_{k,0}.
\end{equation}
Thus, eigenvalue associated with the uniform state around the ring ($k=0$) is split off from the rest even in the limit of $N\rightarrow \infty$, when the other eigenvalues associated with $k\neq0$ close up. When $\beta \rightarrow 0$ and $g \rightarrow 0$, the uniform state is at the top of the band. With the asymmetric advection term $g>0$ (corresponding to a clockwise bias), the above spectrum will become complex.
For a later use, here we briefly consider a limit of $g=0$ and $\beta \rightarrow 0$. In this case, the eigenvalue corresponding to the uniform state ($k=0$) is $2-\beta N$.
The eigenvalue is the largest and at the top of the band when $\beta=0$. The second largest eigenvalue with $k=\pm 2\pi /N$ is $2\cos (2 \pi/N) \approx 2\big(1-\frac{1}{2} (2\pi/N)^2 + O(N^{-4})\big) \approx 2 - \frac{2\pi^2}{N^2}$.
Thus, for the uniform state to be at the top of the spectrum, we require $\beta < \frac{2\pi^2}{N^3}$.
\begin{figure} [!htb]
\centering
\includegraphics[clip,width=1.0\columnwidth]{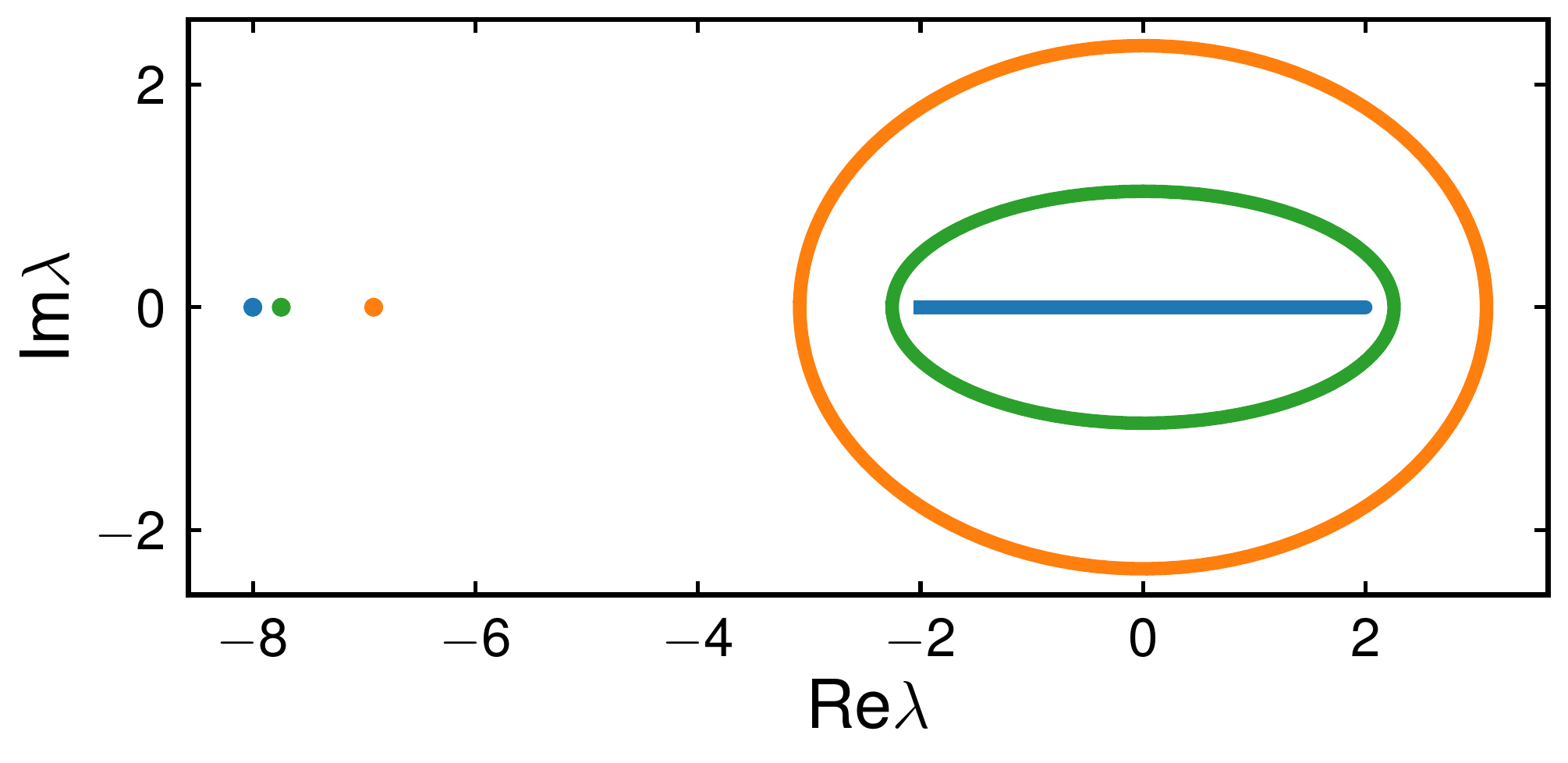}
\caption{
Complex eigenvalue spectrum of the discrete no-disorder model with $N=500$, $\beta=10/N$ and various $g=0,~0.5,~1.0$.
The detached eigenvalues \big(given by $2\cosh(g) - \beta N$\big) correspond to uniform states ($k=0$). 
}
\end{figure}
\subsection{Calculation of eigenvalue spectrum with single anomalous hopping matrix element using a Green's function method}
\begin{figure} [!htb]
\centering
\includegraphics[clip,width=1.0\columnwidth]{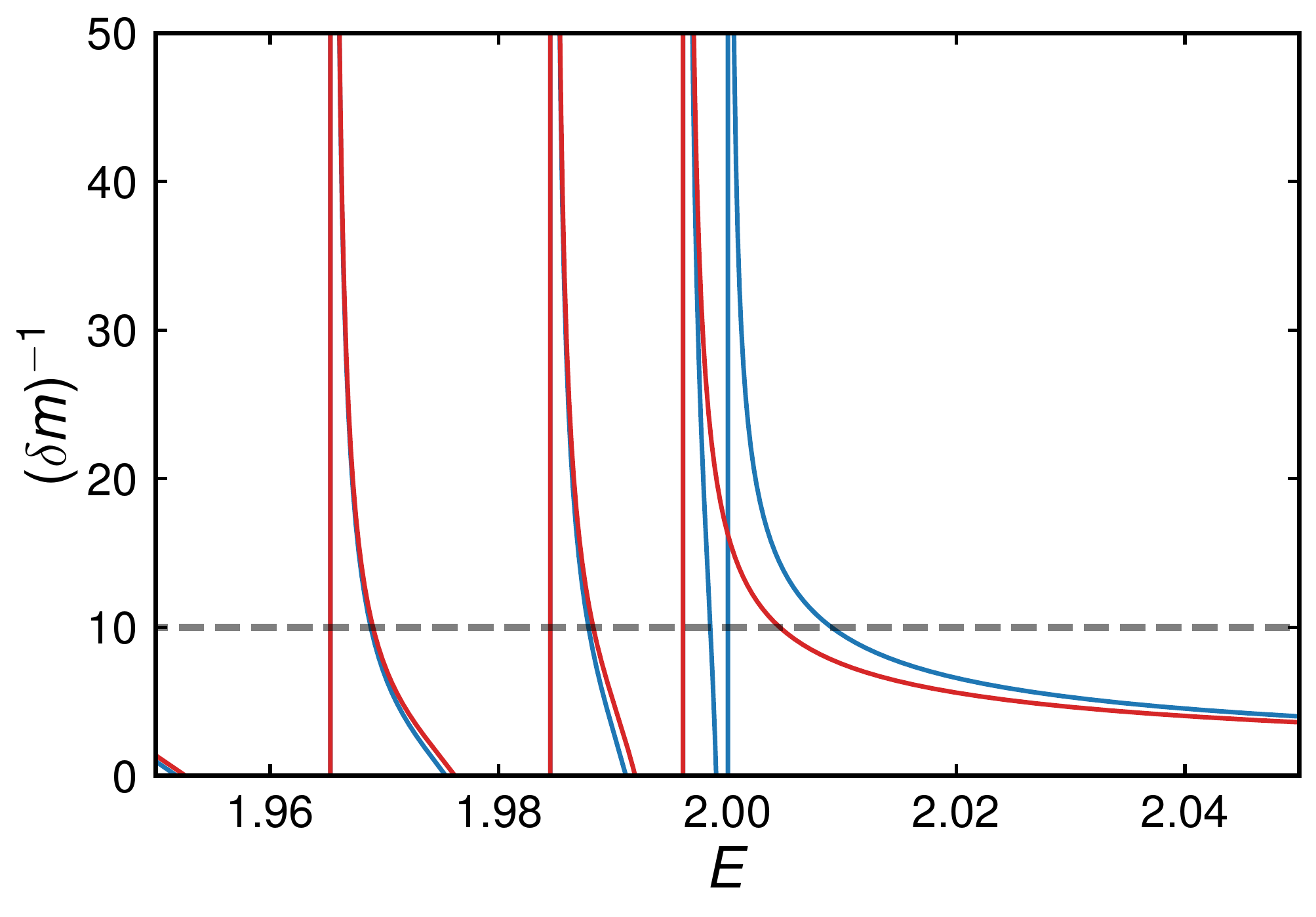}
\caption{
Graphical solution of Eq.~(\ref{C20}) for the shifted eigenvalue spectrum $E$ for a ring with $N=101$ and a single impurity for a particular value of $1/\delta m$ (horizontal dashed line).  
Blue lines represent analytically obtained eigenvalues as a function of $(\delta m)^{-1}$ with $\beta=0$ (no global inhibition), and red lines are with $\beta=1$ (strong global inhibition).
The largest eigenvalue breaks off from what becomes the band of extended states when $N\rightarrow \infty$ and the eigenvalues close up ($E<2$), and corresponds to a localized eigenvector.
We also confirmed that numerical solution agrees well with the analytical solution presented here.
}
\label{Gsol}
\end{figure}

Here, we consider how perturbation to a pair of hopping terms $\delta{\boldsymbol{M}} = \delta m (\ket{1}\bra{2} + \ket{2}\bra{1})$ shifts the eigenvalue spectrum of the unperturbed operator $\boldsymbol{M} = \sum_{l,l'=1}^{N} [\delta_{l,l'+1} + \delta_{l+1,l'} - \beta] \ket{l}\bra{l'} $ following the Green function method presented in \cite{ziman1972principles}.
In a matrix notation, $\boldsymbol{M} + \delta\boldsymbol{M}$ for the case $N=5$ is given by
\begin{equation}
\begin{split}
  \begin{pmatrix}
    -\beta & 1-\beta & -\beta & -\beta & 1-\beta \\
    1-\beta & -\beta & 1-\beta & -\beta & -\beta \\
    -\beta & 1-\beta & -\beta & 1-\beta & -\beta \\
    -\beta & -\beta & 1-\beta & -\beta & 1-\beta \\
    1-\beta & -\beta & -\beta & 1-\beta & -\beta 
  \end{pmatrix}
  +
    \begin{pmatrix}
    0 & \delta m & 0 & 0 & 0 \\
    \delta m & 0 & 0 & 0 & 0 \\
    0 & 0 & 0 & 0 & 0 \\
    0 & 0 & 0 & 0 & 0 \\
    0 & 0 & 0 & 0 & 0
  \end{pmatrix}.
\end{split}
\end{equation}
As discussed in the previous section, $\boldsymbol{M}$ is a circulant matrix in real space, and can be diagonalized by the basis set $\{ \tilde{\ket{k}}\}$,
\begin{equation}
\begin{split}
\boldsymbol{M}=\sum_{k} (2\cos k-\beta N \delta_{k,0}) \tilde{\ket{k}}\tilde{\bra{k}}
\equiv \sum_{k} E(k) \tilde{\ket{k}}\tilde{\bra{k}}.
\end{split}
\end{equation}
Here, we define a linear operator $\boldsymbol{A}$ as,
\begin{equation}
\boldsymbol{A} (E) = E \boldsymbol{I} - \boldsymbol{M},
\end{equation}
where $\boldsymbol{I}$ is the identity matrix,
\begin{equation}
\begin{split}
\boldsymbol{I}&=\sum_{l} \ket{l}\bra{l}=\sum_{k} \tilde{\ket{k}}\tilde{\bra{k}}.
\end{split}
\end{equation}
Clearly, the roots of the equation $0 = \boldsymbol{A}(E) \tilde{\ket{k}}$ with the unperturbed matrix $\boldsymbol{M}$ given the eigenvalue spectrum $E(k)$,
\begin{equation}
0 = \boldsymbol{A}(E) \tilde{\ket{k}} = (E \boldsymbol{I} - \boldsymbol{M}) \tilde{\ket{k}} = (E - E(k)) \tilde{\ket{k}}.
\end{equation}
Now consider a perturbation $\boldsymbol{M} \rightarrow \boldsymbol{M} + \delta\boldsymbol{M} $. The shifted eigenvalues $E$ can be obtained by solving
\begin{equation}
(\boldsymbol{A}(E) - \delta\boldsymbol{M}) \ket{\psi}= 0.
\end{equation}
The inverse of $\boldsymbol{A}$ is obtained as,
\begin{equation}
\begin{split}
\boldsymbol{A}^{-1} &= \sum_{k} \frac{1}{E-E(k)} \tilde{\ket{k}}\tilde{\bra{k}}\\
&= \sum_{l,l'} \frac{1}{N} \sum_{k} \frac{e^{ik(l-l')} }{E-E(k)} \ket{l} \bra{l'},
\end{split}
\end{equation}
An equivalent relation reads,
\begin{equation}
\boldsymbol{A}^{-1} ( \boldsymbol{A} - \delta\boldsymbol{M}) \ket{\psi}
= ( 1 - \boldsymbol{A}^{-1}\delta\boldsymbol{M}) \ket{\psi}= 0.
\end{equation}
Next, we substitute $\delta \boldsymbol{M} = \delta m (\ket{1}\bra{2} + \ket{2}\bra{1})$ and find
\begin{equation}
\begin{split}
&\boldsymbol{A}^{-1} \delta \boldsymbol{M} = \sum_{l,l'} \frac{1}{N} \sum_{k} \frac{e^{ik(l-l')} }{E-E(k)} \ket{l} \bra{l'} \big(\delta m (\ket{1}\bra{2} + \ket{2}\bra{1})\big)\\
&= \sum_{l} \frac{1}{N} \sum_{k} \frac{ \delta m e^{ikl} }{E-E(k)} \ket{l} \bra{2} + 
\sum_{l} \frac{1}{N} \sum_{k} \frac{ \delta m e^{ik(l-1)} }{E-E(k)} \ket{l} \bra{1}.
\end{split}
\end{equation}
The following four matrix elements,
\begin{equation}
\begin{split}
&\bra{1} \boldsymbol{A}^{-1} \delta \boldsymbol{M} \ket{1} = \frac{1}{N} \sum_{k} \frac{ \delta m e^{-ik} }{E-E(k)},\\
&\bra{2}\boldsymbol{A}^{-1} \delta \boldsymbol{M} \ket{1} = \sum_{l} \frac{1}{N} \sum_{k} \frac{ \delta m }{E-E(k)}\\
&\bra{1}\boldsymbol{A}^{-1} \delta \boldsymbol{M} \ket{2} =\frac{1}{N} \sum_{k} \frac{ \delta m }{E-E(k)},\\
&\bra{2}\boldsymbol{A}^{-1} \delta \boldsymbol{M} \ket{2} =\sum_{l} \frac{1}{N} \sum_{k} \frac{ \delta m e^{ik} }{E-E(k)},
\end{split}
\end{equation}
lead to two coupled linear equations,
\begin{equation}
\begin{split}
0 = \psi_0 - \bra{1} \boldsymbol{A}^{-1} \delta \boldsymbol{M} \ket{1} \psi_0 - \bra{1} \boldsymbol{A}^{-1} \delta \boldsymbol{M} \ket{2} \psi_1 \\
0 = \psi_1 - \bra{2} \boldsymbol{A}^{-1} \delta \boldsymbol{M} \ket{2} \psi_1 - \bra{2} \boldsymbol{A}^{-1} \delta \boldsymbol{M} \ket{1} \psi_0
\end{split}
\end{equation}
Since $\bra{1} \boldsymbol{A}^{-1} \delta \boldsymbol{M} \ket{1} = \bra{2} \boldsymbol{A}^{-1} \delta \boldsymbol{M} \ket{2}$ and $\bra{1} \boldsymbol{A}^{-1} \delta \boldsymbol{M} \ket{2} = \bra{2} \boldsymbol{A}^{-1} \delta \boldsymbol{M} \ket{1}$, we can infer $\psi_0 = \psi_1$, and find that the shifted eigenvalues $E$ are the roots of,
\begin{equation}
\begin{split}
\frac{1}{\delta m} &= \frac{1}{N} \sum_{k} \frac{ 1 + e^{ik} }{E-(2\cos k-\beta N \delta_{k,0})},\\
k&=\frac{2\pi s}{N}, s=0,\pm1,...
\end{split}
\label{C20}
\end{equation}
Given $\delta m$, solutions of the above equation can be obtained graphically as in Fig.~\ref{Gsol}.

\subsection{Exact solution: A single imperfection in continuous LEGI model}
\begin{figure} [!htb]
\centering
\includegraphics[clip,width=0.78\columnwidth]{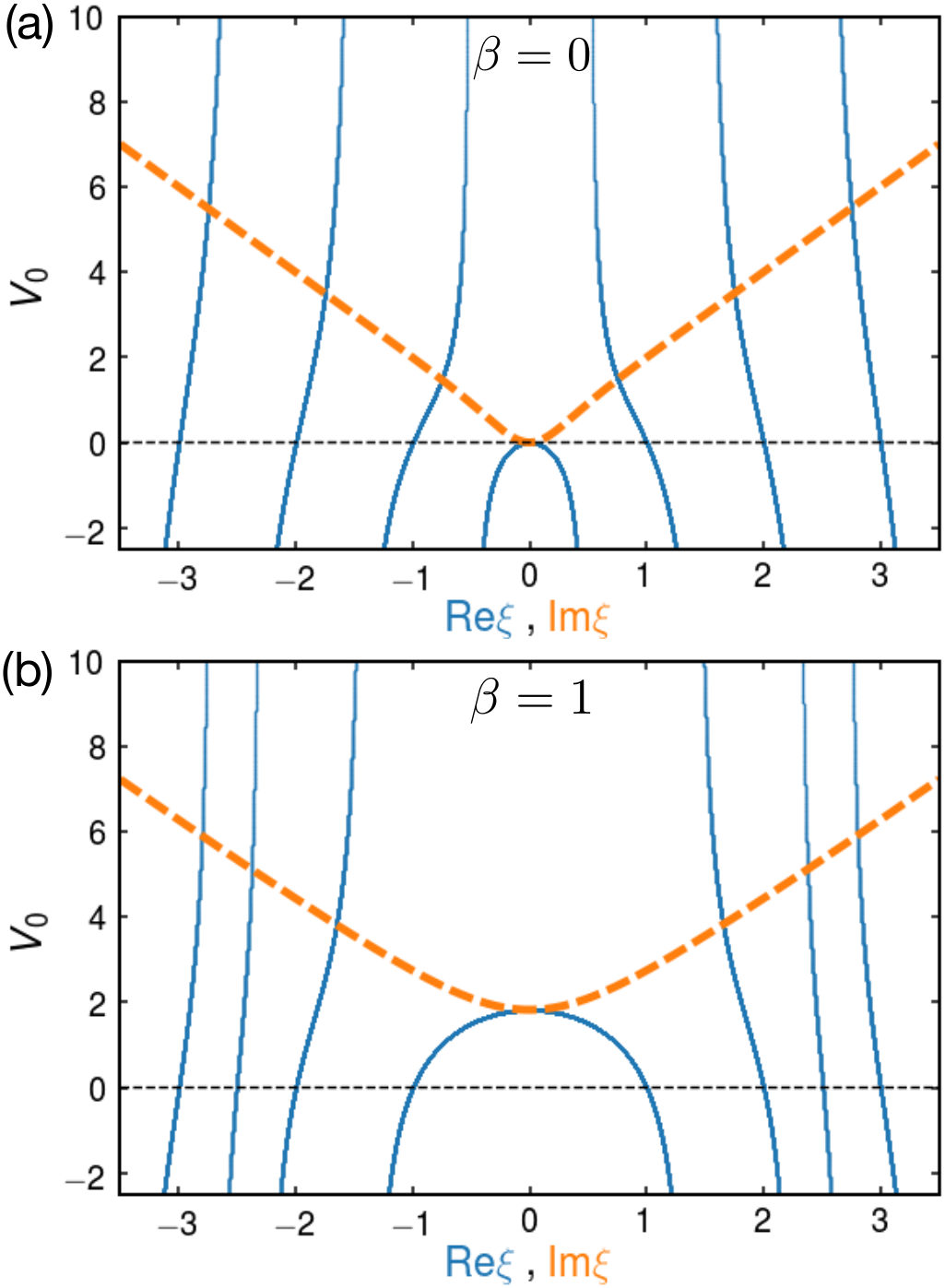}
\caption{
Plot of the derived relation between $\xi$ and $V_0$,
$V_0  = - \frac{2\xi \sin (\xi \pi)}{\cos(\xi \pi) + \beta \frac{2\sin(\xi \pi)}{\xi (\xi^2 - 2\pi \beta)}} \equiv f(\xi)$.
The solid blue lines correspond to $V_0 = f(\xi)$ for real $\xi$, and the dashed orange line corresponds to $V_0 = f(\xi)$ for imaginary $\xi$.
This dashed (orange) line corresponds to the trajectory of localized state and without impurity ($V_0=0$).
(a) Without global inhibition ($\beta=0$), there is a uniform state at the origin ($\xi=0$) with the largest eigenvalue along $V_0=0$, and when $V_0$ becomes finite, $\xi$ immediately becomes imaginary to localize.
(b) With global inhibition ($\beta=1$), the state with $\xi \pm 1$ has the largest eigenvalue reflecting the shift of eigenvalue of the uniform state $\xi =0$. 
}
\label{Fig10}
\end{figure}

Here, we study eigenfunctions $\psi(\theta)$ of the linear operator $\mathcal{L}$,
\begin{equation}
\mathcal{L} \psi(\theta) \equiv \frac{\partial^2 \psi(\theta)}{\partial \theta^2} - \beta \int_0^{2\pi} d \theta \psi(\theta) + V_0 \delta (\theta) \psi(\theta).
\end{equation}
Without the imperfection at $\theta = 0$ represented by the delta-functions, the system has continous rotational symmetry and we can assume eigenfunctions proportional to $e^{i\xi \theta}$.
We decompose $\phi(\theta)$ into Fourier modes according to
\begin{equation}
\psi(\theta) = \sum_{\xi} \widetilde{\psi}(\xi) e^{i\xi \theta},
\end{equation}
\begin{equation}
\widetilde{\psi}(\xi) = \frac{1}{2\pi} \int_0^{2\pi} \psi(\theta) e^{-i\xi\theta} d\theta.
\end{equation}
Without the imperfection ($V_0 = 0$), the eigenvalues $\lambda_{\xi}$ of these extended eigenfunctions are given by
\begin{equation}
(-\xi^2 - 2\pi\beta \delta_{\xi,0}) e^{i\xi \theta} = \lambda_{\xi} e^{i\xi \theta}.
\end{equation}
To understand an imperfection at the origin ($V_0 \neq 0$), we start from an ansatz,
\begin{equation}
\psi(\theta) = \cos(\xi (\theta - \pi)) + C,
\end{equation}
where $\xi$ can be either a real or complex number and $C$ is to be determined.
Note that there can be a jump in the slope of $\psi(\theta=0)$ at the origin with finite $V_0 \delta(\theta)$.\\
\textbf{For $\theta \neq 0, \xi \neq 0$):}\\
If we substitute the above ansatz, we get
\begin{equation}
\mathcal{L} \psi(\theta) = - \xi^2 \bigg[ \cos\big(\xi(\theta -\pi)\big) + \frac{2\beta}{\xi^2} \bigg( \frac{\sin(\xi \pi)}{\xi}+ \pi C\bigg) \bigg].
\end{equation}
Thus, for $\psi(\theta)$ to be an eigenfunction of $\mathcal{L}$, the constant $C$ should satisfy,
\begin{equation}
C = \frac{2\beta \sin(\xi \pi)}{\xi \big(\xi^2 - 2\pi \beta \big)}.\\
\end{equation}
\textbf{For $\theta \neq 0, \xi = 0$):}\\
When $\xi=0$ and the eigenfuction is uniform,
\begin{equation}
\psi(\theta) = 1+C,~\mathcal{L} \psi(\theta) = -2\pi\beta(1+C).
\end{equation}
Note that the eigenvalue of this uniform state $-2 \pi \beta$ is at the top of the band without inhibition ($\beta=0$), but the state drops down in the eigenvalue spectrum with finite $\beta$.\\
\textbf{At the origin $\theta=0$)}\\
Due to the point impurity at the origin there can be a discontinuity in the slope of eigenfunctions at the origin, proportional to the disorder strength $V_0$,
\begin{equation}
\lim_{\epsilon \rightarrow 0} \int_{-\epsilon}^{\epsilon} d\theta \mathcal{L} \psi(\theta) = \psi'(0) - \psi'(2\pi) + V_0 \psi(0) = 0.
\end{equation}
\textbf{At the origin $\theta=0$, $\xi \neq 0$)}\\
By substituting $\psi'(\theta)= -\xi \sin\big(\xi(\theta-\pi)\big)$ into the previous equation we get,
\begin{equation}
\begin{split}
2\xi \sin(\xi \pi) + V_0 \cos(\xi \pi) + V_0 C = 0\\
\Leftrightarrow
C = - \frac{2\xi \sin (\xi \pi)}{V_0} - \cos(\xi \pi)
\end{split}
\end{equation}
By equating the two equations for $C$, 
\begin{equation}
\frac{2\beta \sin(\xi \pi)}{\xi \big(\xi^2 - 2\pi \beta \big)} = - \frac{2\xi \sin (\xi \pi)}{V_0} - \cos(\xi \pi),
\end{equation}
we obtain a relation,
\begin{equation}
V_0 = - \frac{2\xi \sin (\xi \pi)}{\cos(\xi \pi) + \beta \frac{2\sin(\xi \pi)}{\xi (\xi^2 - 2\pi \beta)}}.
\end{equation}
Given $\beta$ and $V_0$, we can determine wave numbers $\xi$ as roots of the above equation.
Furthermore, $\xi$ becomes purely imaginary, $\xi = i \kappa$ when $V_0$ is large, which gives a localized mode as,
\begin{equation}
\psi(\theta) = \cosh(\kappa(\theta - \pi)) + C.
\end{equation}
Our results for $\beta=1$ and $\beta=0$ are summarized in Fig.~\ref{Fig10} (a) and (b) respectively.
\begin{figure} [!b]
\centering
\includegraphics[clip,width=0.85\columnwidth]{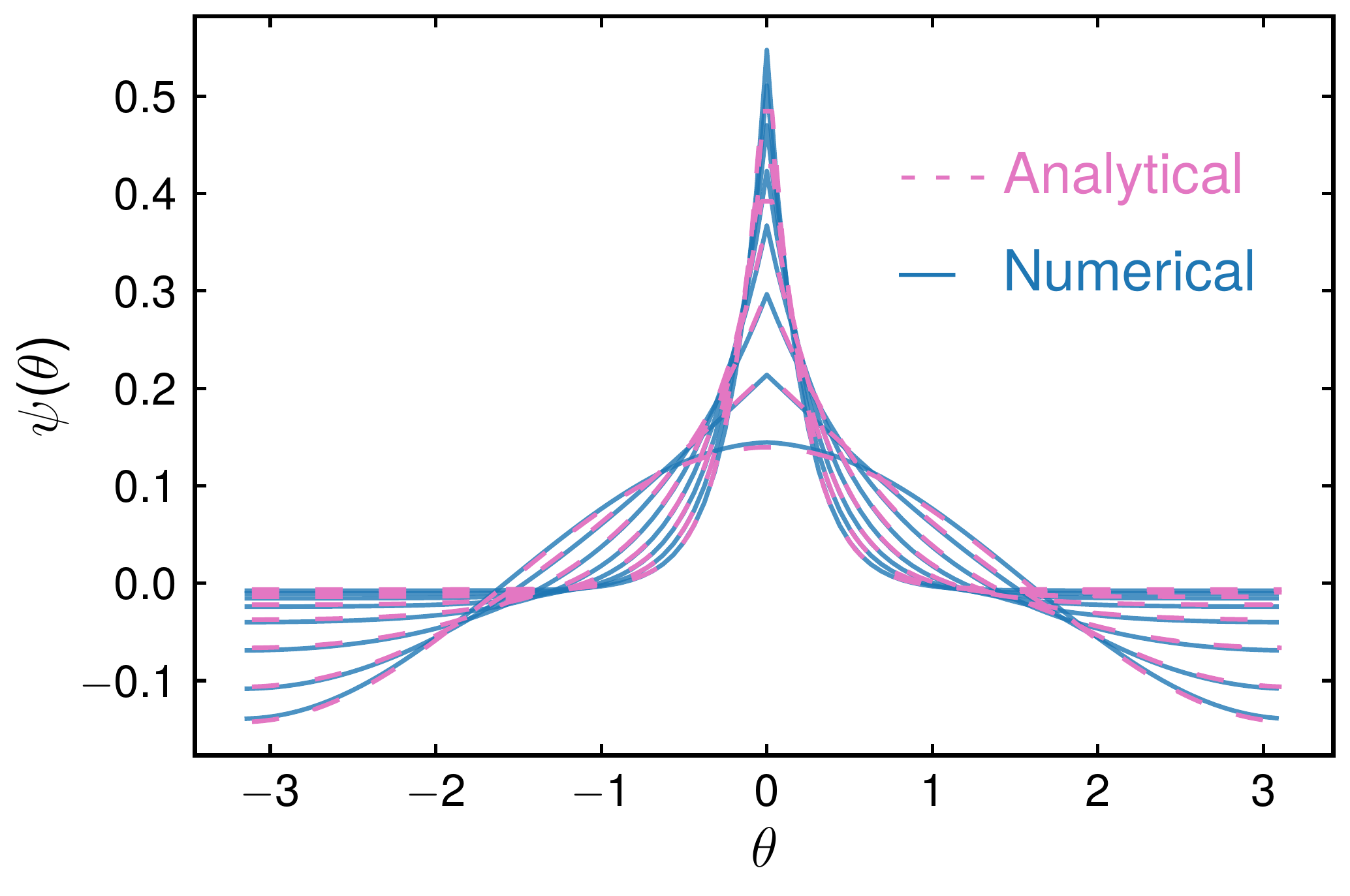}
\caption{
Localization of the principal eigenvalue of the ring model with a single imperfection at the origin.
With a finite global inhibition such that $\beta = 1 \gg \frac{2\pi^2}{N^3}$, the uniform state $k=0$ is no longer at the top of the eigenvalue spectrum.
Instead, a low frequency eigenmode with finite wave number $k\neq0$ will localize as we increase the single site perturbation.
Pink dashed lines correspond to the analytical solutions, while the blue solid lines are eigenvectors obtained with numerical diagonalization.
}
\end{figure}

\clearpage
\bibliography{reference}

\begin{thebibliography}{35}%
\makeatletter
\providecommand \@ifxundefined [1]{%
 \@ifx{#1\undefined}
}%
\providecommand \@ifnum [1]{%
 \ifnum #1\expandafter \@firstoftwo
 \else \expandafter \@secondoftwo
 \fi
}%
\providecommand \@ifx [1]{%
 \ifx #1\expandafter \@firstoftwo
 \else \expandafter \@secondoftwo
 \fi
}%
\providecommand \natexlab [1]{#1}%
\providecommand \enquote  [1]{``#1''}%
\providecommand \bibnamefont  [1]{#1}%
\providecommand \bibfnamefont [1]{#1}%
\providecommand \citenamefont [1]{#1}%
\providecommand \href@noop [0]{\@secondoftwo}%
\providecommand \href [0]{\begingroup \@sanitize@url \@href}%
\providecommand \@href[1]{\@@startlink{#1}\@@href}%
\providecommand \@@href[1]{\endgroup#1\@@endlink}%
\providecommand \@sanitize@url [0]{\catcode `\\12\catcode `\$12\catcode
  `\&12\catcode `\#12\catcode `\^12\catcode `\_12\catcode `\%12\relax}%
\providecommand \@@startlink[1]{}%
\providecommand \@@endlink[0]{}%
\providecommand \url  [0]{\begingroup\@sanitize@url \@url }%
\providecommand \@url [1]{\endgroup\@href {#1}{\urlprefix }}%
\providecommand \urlprefix  [0]{URL }%
\providecommand \Eprint [0]{\href }%
\providecommand \doibase [0]{http://dx.doi.org/}%
\providecommand \selectlanguage [0]{\@gobble}%
\providecommand \bibinfo  [0]{\@secondoftwo}%
\providecommand \bibfield  [0]{\@secondoftwo}%
\providecommand \translation [1]{[#1]}%
\providecommand \BibitemOpen [0]{}%
\providecommand \bibitemStop [0]{}%
\providecommand \bibitemNoStop [0]{.\EOS\space}%
\providecommand \EOS [0]{\spacefactor3000\relax}%
\providecommand \BibitemShut  [1]{\csname bibitem#1\endcsname}%
\let\auto@bib@innerbib\@empty
\bibitem [{\citenamefont {Seung}(2012)}]{seung2012connectome}%
  \BibitemOpen
  \bibfield  {author} {\bibinfo {author} {\bibfnamefont {S.}~\bibnamefont
  {Seung}},\ }\href@noop {} {\emph {\bibinfo {title} {Connectome: how the
  brain's wiring makes us who we are}}}\ (\bibinfo  {publisher} {Houghton
  Mifflin Harcourt},\ \bibinfo {year} {2012})\BibitemShut {NoStop}%
\bibitem [{\citenamefont {Sporns}\ \emph {et~al.}(2005)\citenamefont {Sporns},
  \citenamefont {Tononi},\ and\ \citenamefont {K{\"o}tter}}]{sporns2005human}%
  \BibitemOpen
  \bibfield  {author} {\bibinfo {author} {\bibfnamefont {O.}~\bibnamefont
  {Sporns}}, \bibinfo {author} {\bibfnamefont {G.}~\bibnamefont {Tononi}}, \
  and\ \bibinfo {author} {\bibfnamefont {R.}~\bibnamefont {K{\"o}tter}},\
  }\href@noop {} {\bibfield  {journal} {\bibinfo  {journal} {PLoS computational
  biology}\ }\textbf {\bibinfo {volume} {1}},\ \bibinfo {pages} {e42} (\bibinfo
  {year} {2005})}\BibitemShut {NoStop}%
\bibitem [{\citenamefont {Sompolinsky}\ \emph {et~al.}(1988)\citenamefont
  {Sompolinsky}, \citenamefont {Crisanti},\ and\ \citenamefont
  {Sommers}}]{sompolinsky1988chaos}%
  \BibitemOpen
  \bibfield  {author} {\bibinfo {author} {\bibfnamefont {H.}~\bibnamefont
  {Sompolinsky}}, \bibinfo {author} {\bibfnamefont {A.}~\bibnamefont
  {Crisanti}}, \ and\ \bibinfo {author} {\bibfnamefont {H.-J.}\ \bibnamefont
  {Sommers}},\ }\href@noop {} {\bibfield  {journal} {\bibinfo  {journal}
  {Physical review letters}\ }\textbf {\bibinfo {volume} {61}},\ \bibinfo
  {pages} {259} (\bibinfo {year} {1988})}\BibitemShut {NoStop}%
\bibitem [{\citenamefont {Rajan}\ and\ \citenamefont
  {Abbott}(2006)}]{rajan2006eigenvalue}%
  \BibitemOpen
  \bibfield  {author} {\bibinfo {author} {\bibfnamefont {K.}~\bibnamefont
  {Rajan}}\ and\ \bibinfo {author} {\bibfnamefont {L.}~\bibnamefont {Abbott}},\
  }\href@noop {} {\bibfield  {journal} {\bibinfo  {journal} {Physical review
  letters}\ }\textbf {\bibinfo {volume} {97}},\ \bibinfo {pages} {188104}
  (\bibinfo {year} {2006})}\BibitemShut {NoStop}%
\bibitem [{\citenamefont {Ben-Yishai}\ \emph {et~al.}(1995)\citenamefont
  {Ben-Yishai}, \citenamefont {Bar-Or},\ and\ \citenamefont
  {Sompolinsky}}]{ben1995theory}%
  \BibitemOpen
  \bibfield  {author} {\bibinfo {author} {\bibfnamefont {R.}~\bibnamefont
  {Ben-Yishai}}, \bibinfo {author} {\bibfnamefont {R.~L.}\ \bibnamefont
  {Bar-Or}}, \ and\ \bibinfo {author} {\bibfnamefont {H.}~\bibnamefont
  {Sompolinsky}},\ }\href@noop {} {\bibfield  {journal} {\bibinfo  {journal}
  {Proceedings of the National Academy of Sciences}\ }\textbf {\bibinfo
  {volume} {92}},\ \bibinfo {pages} {3844} (\bibinfo {year}
  {1995})}\BibitemShut {NoStop}%
\bibitem [{\citenamefont {Hansel}\ and\ \citenamefont
  {Sompolinsky}(1998)}]{hansel199813}%
  \BibitemOpen
  \bibfield  {author} {\bibinfo {author} {\bibfnamefont {D.}~\bibnamefont
  {Hansel}}\ and\ \bibinfo {author} {\bibfnamefont {H.}~\bibnamefont
  {Sompolinsky}},\ }\href@noop {} {\emph {\bibinfo {title} {13 modeling feature
  selectivity in local cortical circuits}}}\ (\bibinfo  {publisher} {MIT
  Press},\ \bibinfo {year} {1998})\BibitemShut {NoStop}%
\bibitem [{\citenamefont {Zhang}(1996)}]{Zhang1996}%
  \BibitemOpen
  \bibfield  {author} {\bibinfo {author} {\bibfnamefont {K.}~\bibnamefont
  {Zhang}},\ }\href {http://www.jneurosci.org/content/jneuro/16/6/2112.full.pdf
  http://www.ncbi.nlm.nih.gov/pubmed/8604055} {\bibfield  {journal} {\bibinfo
  {journal} {The Journal of neuroscience : the official journal of the Society
  for Neuroscience}\ }\textbf {\bibinfo {volume} {16}},\ \bibinfo {pages}
  {2112} (\bibinfo {year} {1996})}\BibitemShut {NoStop}%
\bibitem [{\citenamefont {Knierim}\ and\ \citenamefont
  {Zhang}(2012)}]{knierim2012attractor}%
  \BibitemOpen
  \bibfield  {author} {\bibinfo {author} {\bibfnamefont {J.~J.}\ \bibnamefont
  {Knierim}}\ and\ \bibinfo {author} {\bibfnamefont {K.}~\bibnamefont
  {Zhang}},\ }\href@noop {} {\bibfield  {journal} {\bibinfo  {journal} {Annual
  review of neuroscience}\ }\textbf {\bibinfo {volume} {35}},\ \bibinfo {pages}
  {267} (\bibinfo {year} {2012})}\BibitemShut {NoStop}%
\bibitem [{\citenamefont {Xie}\ \emph {et~al.}(2002)\citenamefont {Xie},
  \citenamefont {Hahnloser},\ and\ \citenamefont {Seung}}]{xie2002double}%
  \BibitemOpen
  \bibfield  {author} {\bibinfo {author} {\bibfnamefont {X.}~\bibnamefont
  {Xie}}, \bibinfo {author} {\bibfnamefont {R.~H.}\ \bibnamefont {Hahnloser}},
  \ and\ \bibinfo {author} {\bibfnamefont {H.~S.}\ \bibnamefont {Seung}},\
  }\href@noop {} {\bibfield  {journal} {\bibinfo  {journal} {Physical Review
  E}\ }\textbf {\bibinfo {volume} {66}},\ \bibinfo {pages} {041902} (\bibinfo
  {year} {2002})}\BibitemShut {NoStop}%
\bibitem [{\citenamefont {Renart}\ \emph {et~al.}(2003)\citenamefont {Renart},
  \citenamefont {Song},\ and\ \citenamefont {Wang}}]{renart2003robust}%
  \BibitemOpen
  \bibfield  {author} {\bibinfo {author} {\bibfnamefont {A.}~\bibnamefont
  {Renart}}, \bibinfo {author} {\bibfnamefont {P.}~\bibnamefont {Song}}, \ and\
  \bibinfo {author} {\bibfnamefont {X.-J.}\ \bibnamefont {Wang}},\ }\href@noop
  {} {\bibfield  {journal} {\bibinfo  {journal} {Neuron}\ }\textbf {\bibinfo
  {volume} {38}},\ \bibinfo {pages} {473} (\bibinfo {year} {2003})}\BibitemShut
  {NoStop}%
\bibitem [{\citenamefont {Kilpatrick}\ and\ \citenamefont
  {Ermentrout}(2013)}]{kilpatrick2013wandering}%
  \BibitemOpen
  \bibfield  {author} {\bibinfo {author} {\bibfnamefont {Z.~P.}\ \bibnamefont
  {Kilpatrick}}\ and\ \bibinfo {author} {\bibfnamefont {B.}~\bibnamefont
  {Ermentrout}},\ }\href@noop {} {\bibfield  {journal} {\bibinfo  {journal}
  {SIAM Journal on Applied Dynamical Systems}\ }\textbf {\bibinfo {volume}
  {12}},\ \bibinfo {pages} {61} (\bibinfo {year} {2013})}\BibitemShut {NoStop}%
\bibitem [{\citenamefont {Itskov}\ \emph {et~al.}(2011)\citenamefont {Itskov},
  \citenamefont {Hansel},\ and\ \citenamefont {Tsodyks}}]{itskov2011short}%
  \BibitemOpen
  \bibfield  {author} {\bibinfo {author} {\bibfnamefont {V.}~\bibnamefont
  {Itskov}}, \bibinfo {author} {\bibfnamefont {D.}~\bibnamefont {Hansel}}, \
  and\ \bibinfo {author} {\bibfnamefont {M.}~\bibnamefont {Tsodyks}},\
  }\href@noop {} {\bibfield  {journal} {\bibinfo  {journal} {Frontiers in
  computational neuroscience}\ }\textbf {\bibinfo {volume} {5}},\ \bibinfo
  {pages} {40} (\bibinfo {year} {2011})}\BibitemShut {NoStop}%
\bibitem [{\citenamefont {Zhong}\ \emph {et~al.}(2018)\citenamefont {Zhong},
  \citenamefont {Lu}, \citenamefont {Schwab},\ and\ \citenamefont
  {Murugan}}]{zhong2018non}%
  \BibitemOpen
  \bibfield  {author} {\bibinfo {author} {\bibfnamefont {W.}~\bibnamefont
  {Zhong}}, \bibinfo {author} {\bibfnamefont {Z.}~\bibnamefont {Lu}}, \bibinfo
  {author} {\bibfnamefont {D.~J.}\ \bibnamefont {Schwab}}, \ and\ \bibinfo
  {author} {\bibfnamefont {A.}~\bibnamefont {Murugan}},\ }\href@noop {}
  {\bibfield  {journal} {\bibinfo  {journal} {arXiv preprint arXiv:1809.11167}\
  } (\bibinfo {year} {2018})}\BibitemShut {NoStop}%
\bibitem [{\citenamefont {Kim}\ \emph {et~al.}(2017)\citenamefont {Kim},
  \citenamefont {Rouault}, \citenamefont {Druckmann},\ and\ \citenamefont
  {Jayaraman}}]{Kim2017}%
  \BibitemOpen
  \bibfield  {author} {\bibinfo {author} {\bibfnamefont {S.~S.}\ \bibnamefont
  {Kim}}, \bibinfo {author} {\bibfnamefont {H.}~\bibnamefont {Rouault}},
  \bibinfo {author} {\bibfnamefont {S.}~\bibnamefont {Druckmann}}, \ and\
  \bibinfo {author} {\bibfnamefont {V.}~\bibnamefont {Jayaraman}},\ }\href
  {\doibase 10.1126/science.aal4835} {\bibfield  {journal} {\bibinfo  {journal}
  {Science}\ }\textbf {\bibinfo {volume} {356}},\ \bibinfo {pages} {849}
  (\bibinfo {year} {2017})}\BibitemShut {NoStop}%
\bibitem [{\citenamefont {Seelig}\ and\ \citenamefont
  {Jayaraman}(2015)}]{Seelig}%
  \BibitemOpen
  \bibfield  {author} {\bibinfo {author} {\bibfnamefont {J.~D.}\ \bibnamefont
  {Seelig}}\ and\ \bibinfo {author} {\bibfnamefont {V.}~\bibnamefont
  {Jayaraman}},\ }\href {\doibase 10.1038/nature14446} {\bibfield  {journal}
  {\bibinfo  {journal} {Nature}\ }\textbf {\bibinfo {volume} {521}},\ \bibinfo
  {pages} {186} (\bibinfo {year} {2015})}\BibitemShut {NoStop}%
\bibitem [{\citenamefont {Green}\ \emph {et~al.}(2017)\citenamefont {Green},
  \citenamefont {Adachi}, \citenamefont {Shah}, \citenamefont {Hirokawa},
  \citenamefont {Magani},\ and\ \citenamefont {Maimon}}]{green2017neural}%
  \BibitemOpen
  \bibfield  {author} {\bibinfo {author} {\bibfnamefont {J.}~\bibnamefont
  {Green}}, \bibinfo {author} {\bibfnamefont {A.}~\bibnamefont {Adachi}},
  \bibinfo {author} {\bibfnamefont {K.~K.}\ \bibnamefont {Shah}}, \bibinfo
  {author} {\bibfnamefont {J.~D.}\ \bibnamefont {Hirokawa}}, \bibinfo {author}
  {\bibfnamefont {P.~S.}\ \bibnamefont {Magani}}, \ and\ \bibinfo {author}
  {\bibfnamefont {G.}~\bibnamefont {Maimon}},\ }\href@noop {} {\bibfield
  {journal} {\bibinfo  {journal} {Nature}\ }\textbf {\bibinfo {volume} {546}},\
  \bibinfo {pages} {101} (\bibinfo {year} {2017})}\BibitemShut {NoStop}%
\bibitem [{\citenamefont {Turner-Evans}\ \emph
  {et~al.}(2017{\natexlab{a}})\citenamefont {Turner-Evans}, \citenamefont
  {Wegener}, \citenamefont {Rouault}, \citenamefont {Franconville},
  \citenamefont {Wolff}, \citenamefont {Seelig}, \citenamefont {Druckmann},\
  and\ \citenamefont {Jayaraman}}]{turner2017angular}%
  \BibitemOpen
  \bibfield  {author} {\bibinfo {author} {\bibfnamefont {D.}~\bibnamefont
  {Turner-Evans}}, \bibinfo {author} {\bibfnamefont {S.}~\bibnamefont
  {Wegener}}, \bibinfo {author} {\bibfnamefont {H.}~\bibnamefont {Rouault}},
  \bibinfo {author} {\bibfnamefont {R.}~\bibnamefont {Franconville}}, \bibinfo
  {author} {\bibfnamefont {T.}~\bibnamefont {Wolff}}, \bibinfo {author}
  {\bibfnamefont {J.~D.}\ \bibnamefont {Seelig}}, \bibinfo {author}
  {\bibfnamefont {S.}~\bibnamefont {Druckmann}}, \ and\ \bibinfo {author}
  {\bibfnamefont {V.}~\bibnamefont {Jayaraman}},\ }\href@noop {} {\bibfield
  {journal} {\bibinfo  {journal} {Elife}\ }\textbf {\bibinfo {volume} {6}}
  (\bibinfo {year} {2017}{\natexlab{a}})}\BibitemShut {NoStop}%
\bibitem [{\citenamefont {Anderson}(1958)}]{anderson1958absence}%
  \BibitemOpen
  \bibfield  {author} {\bibinfo {author} {\bibfnamefont {P.~W.}\ \bibnamefont
  {Anderson}},\ }\href@noop {} {\bibfield  {journal} {\bibinfo  {journal}
  {Physical review}\ }\textbf {\bibinfo {volume} {109}},\ \bibinfo {pages}
  {1492} (\bibinfo {year} {1958})}\BibitemShut {NoStop}%
\bibitem [{\citenamefont {Mott}\ and\ \citenamefont
  {Twose}(1995)}]{mott1995theory}%
  \BibitemOpen
  \bibfield  {author} {\bibinfo {author} {\bibfnamefont {N.~F.}\ \bibnamefont
  {Mott}}\ and\ \bibinfo {author} {\bibfnamefont {W.}~\bibnamefont {Twose}},\
  }in\ \href@noop {} {\emph {\bibinfo {booktitle} {Sir Nevill Mott--65 Years In
  Physics}}}\ (\bibinfo  {publisher} {World Scientific},\ \bibinfo {year}
  {1995})\ pp.\ \bibinfo {pages} {259--316}\BibitemShut {NoStop}%
\bibitem [{\citenamefont {Lee}\ and\ \citenamefont
  {Ramakrishnan}(1985)}]{lee1985disordered}%
  \BibitemOpen
  \bibfield  {author} {\bibinfo {author} {\bibfnamefont {P.~A.}\ \bibnamefont
  {Lee}}\ and\ \bibinfo {author} {\bibfnamefont {T.}~\bibnamefont
  {Ramakrishnan}},\ }\href@noop {} {\bibfield  {journal} {\bibinfo  {journal}
  {Reviews of Modern Physics}\ }\textbf {\bibinfo {volume} {57}},\ \bibinfo
  {pages} {287} (\bibinfo {year} {1985})}\BibitemShut {NoStop}%
\bibitem [{\citenamefont {Roati}\ \emph {et~al.}(2008)\citenamefont {Roati},
  \citenamefont {D'Errico}, \citenamefont {Fallani}, \citenamefont {Fattori},
  \citenamefont {Fort}, \citenamefont {Zaccanti}, \citenamefont {Modugno},
  \citenamefont {Modugno},\ and\ \citenamefont {Inguscio}}]{roati2008anderson}%
  \BibitemOpen
  \bibfield  {author} {\bibinfo {author} {\bibfnamefont {G.}~\bibnamefont
  {Roati}}, \bibinfo {author} {\bibfnamefont {C.}~\bibnamefont {D'Errico}},
  \bibinfo {author} {\bibfnamefont {L.}~\bibnamefont {Fallani}}, \bibinfo
  {author} {\bibfnamefont {M.}~\bibnamefont {Fattori}}, \bibinfo {author}
  {\bibfnamefont {C.}~\bibnamefont {Fort}}, \bibinfo {author} {\bibfnamefont
  {M.}~\bibnamefont {Zaccanti}}, \bibinfo {author} {\bibfnamefont
  {G.}~\bibnamefont {Modugno}}, \bibinfo {author} {\bibfnamefont
  {M.}~\bibnamefont {Modugno}}, \ and\ \bibinfo {author} {\bibfnamefont
  {M.}~\bibnamefont {Inguscio}},\ }\href@noop {} {\bibfield  {journal}
  {\bibinfo  {journal} {Nature}\ }\textbf {\bibinfo {volume} {453}},\ \bibinfo
  {pages} {895} (\bibinfo {year} {2008})}\BibitemShut {NoStop}%
\bibitem [{\citenamefont {Billy}\ \emph {et~al.}(2008)\citenamefont {Billy},
  \citenamefont {Josse}, \citenamefont {Zuo}, \citenamefont {Bernard},
  \citenamefont {Hambrecht}, \citenamefont {Lugan}, \citenamefont
  {Cl{\'e}ment}, \citenamefont {Sanchez-Palencia}, \citenamefont {Bouyer},\
  and\ \citenamefont {Aspect}}]{billy2008direct}%
  \BibitemOpen
  \bibfield  {author} {\bibinfo {author} {\bibfnamefont {J.}~\bibnamefont
  {Billy}}, \bibinfo {author} {\bibfnamefont {V.}~\bibnamefont {Josse}},
  \bibinfo {author} {\bibfnamefont {Z.}~\bibnamefont {Zuo}}, \bibinfo {author}
  {\bibfnamefont {A.}~\bibnamefont {Bernard}}, \bibinfo {author} {\bibfnamefont
  {B.}~\bibnamefont {Hambrecht}}, \bibinfo {author} {\bibfnamefont
  {P.}~\bibnamefont {Lugan}}, \bibinfo {author} {\bibfnamefont
  {D.}~\bibnamefont {Cl{\'e}ment}}, \bibinfo {author} {\bibfnamefont
  {L.}~\bibnamefont {Sanchez-Palencia}}, \bibinfo {author} {\bibfnamefont
  {P.}~\bibnamefont {Bouyer}}, \ and\ \bibinfo {author} {\bibfnamefont
  {A.}~\bibnamefont {Aspect}},\ }\href@noop {} {\bibfield  {journal} {\bibinfo
  {journal} {Nature}\ }\textbf {\bibinfo {volume} {453}},\ \bibinfo {pages}
  {891} (\bibinfo {year} {2008})}\BibitemShut {NoStop}%
\bibitem [{\citenamefont {Segev}\ \emph {et~al.}(2013)\citenamefont {Segev},
  \citenamefont {Silberberg},\ and\ \citenamefont
  {Christodoulides}}]{segev2013anderson}%
  \BibitemOpen
  \bibfield  {author} {\bibinfo {author} {\bibfnamefont {M.}~\bibnamefont
  {Segev}}, \bibinfo {author} {\bibfnamefont {Y.}~\bibnamefont {Silberberg}}, \
  and\ \bibinfo {author} {\bibfnamefont {D.~N.}\ \bibnamefont
  {Christodoulides}},\ }\href@noop {} {\bibfield  {journal} {\bibinfo
  {journal} {Nature Photonics}\ }\textbf {\bibinfo {volume} {7}},\ \bibinfo
  {pages} {197} (\bibinfo {year} {2013})}\BibitemShut {NoStop}%
\bibitem [{\citenamefont {Dyson}(1953)}]{dyson1953dynamics}%
  \BibitemOpen
  \bibfield  {author} {\bibinfo {author} {\bibfnamefont {F.~J.}\ \bibnamefont
  {Dyson}},\ }\href@noop {} {\bibfield  {journal} {\bibinfo  {journal}
  {Physical Review}\ }\textbf {\bibinfo {volume} {92}},\ \bibinfo {pages}
  {1331} (\bibinfo {year} {1953})}\BibitemShut {NoStop}%
\bibitem [{\citenamefont {Ishii}(1973)}]{ishii1973localization}%
  \BibitemOpen
  \bibfield  {author} {\bibinfo {author} {\bibfnamefont {K.}~\bibnamefont
  {Ishii}},\ }\href@noop {} {\bibfield  {journal} {\bibinfo  {journal}
  {Progress of Theoretical Physics Supplement}\ }\textbf {\bibinfo {volume}
  {53}},\ \bibinfo {pages} {77} (\bibinfo {year} {1973})}\BibitemShut {NoStop}%
\bibitem [{\citenamefont {Chaudhuri}\ \emph {et~al.}(2014)\citenamefont
  {Chaudhuri}, \citenamefont {Bernacchia},\ and\ \citenamefont
  {Wang}}]{chaudhuri2014diversity}%
  \BibitemOpen
  \bibfield  {author} {\bibinfo {author} {\bibfnamefont {R.}~\bibnamefont
  {Chaudhuri}}, \bibinfo {author} {\bibfnamefont {A.}~\bibnamefont
  {Bernacchia}}, \ and\ \bibinfo {author} {\bibfnamefont {X.-J.}\ \bibnamefont
  {Wang}},\ }\href@noop {} {\bibfield  {journal} {\bibinfo  {journal} {elife}\
  }\textbf {\bibinfo {volume} {3}},\ \bibinfo {pages} {e01239} (\bibinfo {year}
  {2014})}\BibitemShut {NoStop}%
\bibitem [{\citenamefont {Amir}\ \emph {et~al.}(2016)\citenamefont {Amir},
  \citenamefont {Hatano},\ and\ \citenamefont {Nelson}}]{Amir2016}%
  \BibitemOpen
  \bibfield  {author} {\bibinfo {author} {\bibfnamefont {A.}~\bibnamefont
  {Amir}}, \bibinfo {author} {\bibfnamefont {N.}~\bibnamefont {Hatano}}, \ and\
  \bibinfo {author} {\bibfnamefont {D.~R.}\ \bibnamefont {Nelson}},\ }\href
  {\doibase 10.1103/PhysRevE.93.042310} {\bibfield  {journal} {\bibinfo
  {journal} {Physical Review E}\ }\textbf {\bibinfo {volume} {93}},\ \bibinfo
  {pages} {042310} (\bibinfo {year} {2016})}\BibitemShut {NoStop}%
\bibitem [{\citenamefont {Hatano}\ and\ \citenamefont
  {Nelson}(1997)}]{Hatano1997}%
  \BibitemOpen
  \bibfield  {author} {\bibinfo {author} {\bibfnamefont {N.}~\bibnamefont
  {Hatano}}\ and\ \bibinfo {author} {\bibfnamefont {D.~R.}\ \bibnamefont
  {Nelson}},\ }\href {\doibase 10.1103/PhysRevB.56.8651} {\bibfield  {journal}
  {\bibinfo  {journal} {Physical Review B}\ }\textbf {\bibinfo {volume} {56}},\
  \bibinfo {pages} {8651} (\bibinfo {year} {1997})}\BibitemShut {NoStop}%
\bibitem [{\citenamefont {Shnerb}\ and\ \citenamefont
  {Nelson}(1998)}]{Shnerb1998}%
  \BibitemOpen
  \bibfield  {author} {\bibinfo {author} {\bibfnamefont {N.~M.}\ \bibnamefont
  {Shnerb}}\ and\ \bibinfo {author} {\bibfnamefont {D.~R.}\ \bibnamefont
  {Nelson}},\ }\href {\doibase 10.1103/PhysRevLett.80.5172} {\bibfield
  {journal} {\bibinfo  {journal} {Physical Review Letters}\ }\textbf {\bibinfo
  {volume} {80}},\ \bibinfo {pages} {5172} (\bibinfo {year}
  {1998})}\BibitemShut {NoStop}%
\bibitem [{\citenamefont {Turner-Evans}\ \emph
  {et~al.}(2017{\natexlab{b}})\citenamefont {Turner-Evans}, \citenamefont
  {Wegener}, \citenamefont {Rouault}, \citenamefont {Franconville},
  \citenamefont {Wolff}, \citenamefont {Seelig}, \citenamefont {Druckmann},\
  and\ \citenamefont {Jayaraman}}]{Turner-Evans2017}%
  \BibitemOpen
  \bibfield  {author} {\bibinfo {author} {\bibfnamefont {D.}~\bibnamefont
  {Turner-Evans}}, \bibinfo {author} {\bibfnamefont {S.}~\bibnamefont
  {Wegener}}, \bibinfo {author} {\bibfnamefont {H.}~\bibnamefont {Rouault}},
  \bibinfo {author} {\bibfnamefont {R.}~\bibnamefont {Franconville}}, \bibinfo
  {author} {\bibfnamefont {T.}~\bibnamefont {Wolff}}, \bibinfo {author}
  {\bibfnamefont {J.~D.}\ \bibnamefont {Seelig}}, \bibinfo {author}
  {\bibfnamefont {S.}~\bibnamefont {Druckmann}}, \ and\ \bibinfo {author}
  {\bibfnamefont {V.}~\bibnamefont {Jayaraman}},\ }\href {\doibase
  10.7554/eLife.23496} {\bibfield  {journal} {\bibinfo  {journal} {eLife}\
  }\textbf {\bibinfo {volume} {6}} (\bibinfo {year} {2017}{\natexlab{b}}),\
  10.7554/eLife.23496}\BibitemShut {NoStop}%
\bibitem [{\citenamefont {Heinze}(2017)}]{heinze2017neural}%
  \BibitemOpen
  \bibfield  {author} {\bibinfo {author} {\bibfnamefont {S.}~\bibnamefont
  {Heinze}},\ }\href@noop {} {\bibfield  {journal} {\bibinfo  {journal}
  {Current Biology}\ }\textbf {\bibinfo {volume} {27}},\ \bibinfo {pages}
  {R409} (\bibinfo {year} {2017})}\BibitemShut {NoStop}%
\bibitem [{\citenamefont {Mastrogiuseppe}\ and\ \citenamefont
  {Ostojic}(2017)}]{mastrogiuseppe2017linking}%
  \BibitemOpen
  \bibfield  {author} {\bibinfo {author} {\bibfnamefont {F.}~\bibnamefont
  {Mastrogiuseppe}}\ and\ \bibinfo {author} {\bibfnamefont {S.}~\bibnamefont
  {Ostojic}},\ }\href@noop {} {\bibfield  {journal} {\bibinfo  {journal} {arXiv
  preprint arXiv:1711.09672}\ } (\bibinfo {year} {2017})}\BibitemShut {NoStop}%
\bibitem [{\citenamefont {Furstenberg}\ and\ \citenamefont
  {Kesten}(1960)}]{furstenberg1960products}%
  \BibitemOpen
  \bibfield  {author} {\bibinfo {author} {\bibfnamefont {H.}~\bibnamefont
  {Furstenberg}}\ and\ \bibinfo {author} {\bibfnamefont {H.}~\bibnamefont
  {Kesten}},\ }\href@noop {} {\bibfield  {journal} {\bibinfo  {journal} {The
  Annals of Mathematical Statistics}\ }\textbf {\bibinfo {volume} {31}},\
  \bibinfo {pages} {457} (\bibinfo {year} {1960})}\BibitemShut {NoStop}%
\bibitem [{\citenamefont {Skaggs}\ \emph {et~al.}(1995)\citenamefont {Skaggs},
  \citenamefont {Knierim}, \citenamefont {Kudrimoti},\ and\ \citenamefont
  {McNaughton}}]{skaggs1995model}%
  \BibitemOpen
  \bibfield  {author} {\bibinfo {author} {\bibfnamefont {W.~E.}\ \bibnamefont
  {Skaggs}}, \bibinfo {author} {\bibfnamefont {J.~J.}\ \bibnamefont {Knierim}},
  \bibinfo {author} {\bibfnamefont {H.~S.}\ \bibnamefont {Kudrimoti}}, \ and\
  \bibinfo {author} {\bibfnamefont {B.~L.}\ \bibnamefont {McNaughton}},\ }in\
  \href@noop {} {\emph {\bibinfo {booktitle} {Advances in neural information
  processing systems}}}\ (\bibinfo {year} {1995})\ pp.\ \bibinfo {pages}
  {173--180}\BibitemShut {NoStop}%
\bibitem [{\citenamefont {Ziman}(1972)}]{ziman1972principles}%
  \BibitemOpen
  \bibfield  {author} {\bibinfo {author} {\bibfnamefont {J.~M.}\ \bibnamefont
  {Ziman}},\ }\href@noop {} {\emph {\bibinfo {title} {Principles of the Theory
  of Solids}}}\ (\bibinfo  {publisher} {Cambridge university press},\ \bibinfo
  {year} {1972})\BibitemShut {NoStop}%
\end{thebibliography}%

\end{document}